\documentclass[structabstract]{aa}
\usepackage{txfonts}
\usepackage{graphicx}
\usepackage{natbib}
\usepackage{longtable}
\usepackage{subeqnarray}
\usepackage{cases}
\usepackage[colorlinks=true,citecolor=blue]{hyperref}

\begin{document}
\title{Kinetic temperature of massive star-forming molecular clumps measured with formaldehyde}
\subtitle{IV. The ALMA view of N113 and N159W in the LMC}

\author{X. D. Tang\inst{\ref{inst1},\ref{inst2},\ref{inst3}}
\and C. Henkel\inst{\ref{inst3},\ref{inst4},\ref{inst1}}
\and K. M. Menten\inst{\ref{inst3}}
\and Y. Gong\inst{\ref{inst3}}
\and C. -H. R. Chen\inst{\ref{inst3}}
\and D. L. Li\inst{\ref{inst1},\ref{inst2}}
\and M.-Y. Lee\inst{\ref{inst3},\ref{inst19}}
\and J. G. Mangum\inst{\ref{inst5}}
\and Y. P. Ao\inst{\ref{inst6}}
\and S. M\"{u}hle\inst{\ref{inst7}}
\and S. Aalto\inst{\ref{inst8}}
\and S. Garc\'{\i}a-Burillo\inst{\ref{inst9}}
\and S. Mart\'{\i}n\inst{\ref{inst10},\ref{inst11}}
\and S. Viti\inst{\ref{inst22},\ref{inst12}}
\and S. Muller\inst{\ref{inst8}}
\and F. Costagliola\inst{\ref{inst8}}
\and H. Asiri\inst{\ref{inst4}}
\and S. A. Levshakov\inst{\ref{inst13}}
\and M. Spaans\inst{\ref{inst14}}
\and J. Ott\inst{\ref{inst20}}
\and C. M. V. Impellizzeri\inst{\ref{inst5},\ref{inst21}}
\and Y. Fukui\inst{\ref{inst16}}
\and Y. X. He\inst{\ref{inst1},\ref{inst2}}
\and J. Esimbek\inst{\ref{inst1},\ref{inst2}}
\and J. J. Zhou\inst{\ref{inst1},\ref{inst2}}
\and X. W. Zheng\inst{\ref{inst17}}
\and X. Zhao\inst{\ref{inst1},\ref{inst18}}
\and J. S. Li\inst{\ref{inst1},\ref{inst18}}
}

\titlerunning{Kinetic temperatures in N113 and N159W}
\authorrunning{Tang et al.}

\institute{
Xinjiang Astronomical Observatory, Chinese Academy of Sciences, 830011 Urumqi, PR China \label{inst1}\\
\email{tangxindi@xao.ac.cn}
\and Key Laboratory of Radio Astronomy, Chinese Academy of Sciences, 830011 Urumqi, PR China \label{inst2}
\and Max-Planck-Institut f\"{u}r Radioastronomie, Auf dem H\"{u}gel 69, 53121 Bonn, Germany \label{inst3}\\
\email{chenkel@mpifr-bonn.mpg.de}
\and Astronomy Department, King Abdulaziz University, PO Box 80203, 21589 Jeddah, Saudi Arabia \label{inst4}
\and Korea Astronomy and Space Science Institute, 776 Daedeok-daero, Yuseong-gu, Daejeon 34055, Republic of Korea \label{inst19}
\and National Radio Astronomy Observatory, 520 Edgemont Road, Charlottesville, VA 22903, USA \label{inst5}
\and Purple Mountain Observatory, Chinese Academy of Sciences, Nanjing 210008, PR China \label{inst6}
\and Argelander Institut f\"{u}r Astronomie, Universit\"{a}t Bonn, Auf dem H\"{u}gel 71, 53121 Bonn, Germany \label{inst7}
\and Department of Space, Earth and Environment, Chalmers University of Technology, Onsala Space Observatory, SE-43992 Onsala, Sweden \label{inst8}
\and Observatorio de Madrid, OAN-IGN, Alfonso XII, 3, E-28014-Madrid, Spain \label{inst9}
\and European Southern Observatory, Alonso de C\'{o}rdova 3107, Vitacura, Santiago 763-0355, Chile \label{inst10}
\and Joint ALMA Observatory, Alonso de C\'{o}rdova 3107, Vitacura, Santiago 763-0355, Chile \label{inst11}
\and Leiden Observatory, Leiden University, PO Box 9513, 2300 RA Leiden, The Netherlands \label{inst22}
\and Department of Physics and Astronomy, University College London, Gower Street, London, WC1E 6BT, UK \label{inst12}
\and Ioffe Physical-Technical Institute, Polytekhnicheskaya Str. 26, 194021 St. Petersburg, Russia \label{inst13}
\and Kapteyn Astronomical Institute, University of Groningen, PO Box 800, 9700 AV Groningen, The Netherlands \label{inst14}
\and National Radio Astronomy Observatory, P.O. Box O, 1003 Lopezville Road, Socorro, NM 87801, USA  \label{inst20}
\and Joint ALMA Observatory, Alonso de Cordova 3107, Vitacura, Santiago, Chile \label{inst21}
\and Department of Physics, Nagoya University, Furo-cho, Chikusa-ku Nagoya, 464-8601, Japan \label{inst16}
\and School of Astronomy and Space Science, Nanjing University, 210093 Nanjing, PR China \label{inst17}
\and University of Chinese Academy of Sciences, 100080 Beijing, PR China \label{inst18}
}


\abstract
{We mapped the kinetic temperature structure of two massive star-forming regions, N113 and N159W,
in the Large Magellanic Cloud (LMC). We have used $\sim$1\hbox{$\,.\!\!^{\prime\prime}$}6\,($\sim$0.4\,pc)
resolution measurements of the para-H$_2$CO\,$J_{\rm K_ aK_c}$\,=\,3$_{03}$--2$_{02}$, 3$_{22}$--2$_{21}$,
and 3$_{21}$--2$_{20}$ transitions near 218.5\,GHz to constrain RADEX non-LTE models of the physical conditions.
The gas kinetic temperatures derived from the para-H$_2$CO line ratios
3$_{22}$--2$_{21}$/3$_{03}$--2$_{02}$ and 3$_{21}$--2$_{20}$/3$_{03}$--2$_{02}$
range from 28 to 105\,K in N113 and 29 to 68\,K in N159W.
Distributions of the dense gas traced by para-H$_2$CO agree with those of the 1.3\,mm
dust and \emph{Spitzer}\,8.0\,$\mu$m emission, but do not significantly correlate
with the H$\alpha$ emission. The high kinetic temperatures ($T_{\rm kin}$\,$\gtrsim$\,50\,K)
of the dense gas traced by para-H$_2$CO appear to be correlated with
the embedded infrared sources inside the clouds and/or YSOs in the N113 and N159W regions.
The lower temperatures ($T_{\rm kin}$\,$<$\,50\,K) are measured at the outskirts of the H$_2$CO-bearing
distributions of both N113 and N159W. It seems that the kinetic temperatures of the dense gas
traced by para-H$_2$CO are weakly affected by the external sources of the H$\alpha$ emission.
The non-thermal velocity dispersions of para-H$_2$CO are well correlated with the
gas kinetic temperatures in the N113 region, implying that the higher kinetic temperature traced
by para-H$_2$CO is related to turbulence on a $\sim$0.4\,pc scale. The dense gas heating
appears to be dominated by internal star formation activity, radiation, and/or turbulence.
It seems that the mechanism heating the dense gas of the
star-forming regions in the LMC is consistent with that in Galactic massive
star-forming regions located in the Galactic plane.}

 \maketitle

\section{Introduction}
\label{sect:Introduction}
Molecular gas is the fuel for star formation. Increasing evidence indicates
that its physical properties affect star formation rate,
spatial distribution, and essential properties of the next generation of stars,
like elemental composition and initial mass function
(e.g., \citealt{Paumard2006,Kennicutt1998a,Kennicutt1998b,Klessen2007,Papadopoulos2011,Zhang2018,Tang2019}).
Unfortunately, the most readily accessible molecular transitions,
those of CO, suffer from radiative transfer effects such as high optical
depths or subthermal excitation, and are not a reliable probe of these physical conditions.
Furthermore, many molecules tracing the molecular mass in external galaxies
(e.g., \citealt{Gao2004a,Gao2004b,Wu2005,Zhang2014,Henkel2018,Li2021}),
including CO, HCN, HCO$^+$, HNC, and CS, suffer from a coupled sensitivity
to kinetic temperature and density, making them degenerately sensitive
to high spatial densities and low kinetic temperatures or vice versa.
Obtaining information about the individual physical conditions in these
regions requires a molecular tracer that possesses singular sensitivity
to kinetic temperature and that can then be used to also determine,
in a second step, the number density of molecular hydrogen, $n$(H$_2$).

Formaldehyde (H$_2$CO), a slightly asymmetric rotor molecule,
is a ubiquitous molecule in the interstellar medium
(e.g., \citealt{Downes1980,Bieging1982,Henkel1991,Zylka1992,Liszt2006,Mangum2008,Mangum2013,Mangum2019,Ao2013,Tang2013,Tang2014,Ginsburg2015,Ginsburg2016,Guo2016,Yan2019}).
H$_2$CO can be formed on the surface of dust grains by successive hydrogenation of CO
(CO$\rightarrow$HCO$\rightarrow$H$_2$CO) \citep{Watanabe2002,Woon2002,Hidaka2004,Yan2019}.
It is then released into the gas phase by shocks or UV heating and destroyed by photodissociation.
Alternatively, gas phase formation is also possibly contributing to the total
H$_2$CO abundance in Galactic environments such as the dark cloud TMC-1 (e.g., \citealt{Soma2018}).
CH$_3$OH, which is also the product of hydrogenation of CO on grain surfaces
and is hardly formed in the gas phase, is deficient in the the Large Magellanic Cloud (LMC)
(e.g., \citealt{Shimonishi2016}). In the LMC where dust is less abundant than in the Galaxy,
gas phase formation of H$_2$CO may thus be relevant.

H$_2$CO has a stable fractional abundance in star formation regions.
Variations in fractional abundance of H$_2$CO rarely exceed one order
of magnitude at various stages of star formation
(e.g., \citealt{Mangum1990,Mangum1993b,Caselli1993,Johnstone2003,Gerner2014,Tang2017a,Tang2017b,Tang2018b,Zhu2020}).
As further demonstrated in recent studies of formaldehyde in the LMC,
the fractional abundance of para-H$_2$CO in its star-forming regions
has similar values as in Galactic regions of massive star formation
\citep{Tang2017a,Tang2017b,Tang2018b}. H$_2$CO has a rich variety of
millimeter and submillimeter transitions which are a reliable probe to
trace physical conditions of molecular clouds
(e.g., \citealt{Henkel1980,Henkel1983,Mangum1993a,Muhle2007,Mangum2008,Mangum2013,Mangum2019,Ginsburg2011,Tang2017a,Tang2017b,Tang2018a,Tang2018b}).
Since the relative populations of the $K_{\rm a}$ ladders of
H$_2$CO are governed by collisions, line ratios involving
different $K_{\rm a}$ ladders are good tracers of the kinetic
temperature \citep{Mangum1993a}. Therefore, these H$_2$CO line ratios
are an ideal thermometer for tracing the temperature of dense gas
in low-metallicity galaxies with strong UV radiation \citep{Tang2017b}.

The Large Magellanic Cloud is the closest star-forming galaxy to our Milky Way
($\sim$50\,kpc, e.g., \citealt{Pietrzynski2013,Pietrzynski2019}).
In the LMC, the far-ultraviolet (FUV) radiation field is stronger than
in the Milky Way while metallicities are lower \citep{Westerlund1990,Rolleston2002}.
The LMC provides an ideal laboratory for studying star formation, particularly massive
star formation associated with its numerous stellar clusters in such an active low-metallicity galaxy.
Previous observations of CO\,$J$\,=\,1--0 and 3--2 show that the higher gas temperatures
in low/moderate density regions ($n$(H$_2$)\,$\sim$\,a few\,$\times$\,10$^3$\,cm$^{-3}$ in most cases)
are correlated with stronger H$\alpha$ flux in giant molecular clouds of the LMC
\citep{Minamidani2008,Minamidani2011}. Further observations of CO\,$J$\,=\,4--3 to 12--11 indicate that
the gas temperatures are high ($T_{\rm kin}$\,>\,150\,K) at densities $\sim$10$^3$\,cm$^{-3}$
in active star-forming regions of the LMC \citep{Lee2016}. These suggest that FUV heating
may strongly impact on the molecular gas in low density regions ($n$(H$_2$)\,$<$\,10$^5$\,cm$^{-3}$)
of the LMC. In contrast, recent low-resolution (30$''$ or $\sim$7\,pc) observations of
para-H$_2$CO\,$J$\,=\,3--2 show that the kinetic temperatures of the dense gas
($n$(H$_2$)\,$\sim$\,10$^5$\,cm$^{-3}$) are correlated with the ongoing massive
star formation in the LMC \citep{Tang2017b}. This suggests that inside the LMC the gas heating
may be different between molecular cloud layers of lower and higher density.
Although gas heating in low density regions of molecular clouds of the LMC can be
expected to be due to the exciting OB stars of photon dominated regions (PDRs)
(e.g., \citealt{Kaufman1999,Lee2019}), the detailed relationship between the temperature
of the dense gas and the star formation in the LMC is still under debate.

In this paper, we aim to map the kinetic temperature structure
of two massive star-forming regions, N113 and N159W, in the LMC with the
para-H$_2$CO triplet\,($J_{\rm K_aK_c}$\,=\,3$_{03}$--2$_{02}$, 3$_{22}$--2$_{21}$,
and 3$_{21}$--2$_{20}$) near 218.5\,GHz and to investigate the gas heating
process affecting the dense gas. In Sects.\,\ref{sect:Targets-observations-data-reduction}
and \ref{sect:Results}, we introduce our targets, observations of the para-H$_2$CO triplet,
data reduction, and results. We discuss the resulting
kinetic temperatures derived from para-H$_2$CO in Sect.\,\ref{sect:discussion}.
Our main conclusions are summarized in Sect.\,\ref{sect:summary}.
This paper is part of the "Kinetic temperature of massive
star-forming molecular clumps measured with formaldehyde" series of
studies exploring H$_2$CO as a probe of gas conditions in a variety
of Galactic and extragalactic sources.

\section{Targets, observations, and data reduction}
\label{sect:Targets-observations-data-reduction}
\subsection{Targets}
\label{sect:Targets}
The two massive star-forming regions, N113 and N159W, in the LMC were selected from
previous single dish observations with the Atacama Pathfinder EXperiment telescope
(APEX) (beam size $\sim$30$''$) in the course of which three transitions of
para-H$_2$CO\,($J$\,=\,3--2) near $\sim$218.5\,GHz were detected \citep{Tang2017b}.
The dense gas kinetic temperatures of N113 and N159W derived from these
para-H$_2$CO line ratios are 54\,$\pm$\,7\,K and 35\,$\pm$\,4\,K, respectively,
on a scale of $\sim$7\,pc \citep{Tang2017b}.

\begin{table*}[t]
\caption{Observational parameters.}
\label{table:Observational_parameters}
\centering
\begin{tabular}
{cccccccc}
\hline\hline
Source &Central frequency &Bandwidth &Channel width &Beam size &Position angle &RMS\\
&MHz &MHz &km s$^{-1}$ &"$\times$"  &\degr &mJy beam$^{-1}$\\
\hline 
N113  &217104.980  &937.5 &1.35 &1.7$\times$1.3 & 60 &$\sim$2.0\\
      &218475.632  &937.5 &1.34 &1.6$\times$1.3 & 61 &$\sim$2.0\\
      &230538.000  &937.5 &1.27 &1.6$\times$1.2 & 61 &$\sim$2.0\\
\hline
N159W &217104.980  &937.5 &1.35 &1.6$\times$1.3 & 55 &$\sim$2.0\\
      &218475.632  &937.5 &1.34 &1.6$\times$1.3 & 56 &$\sim$2.0\\
      &230538.000  &937.5 &1.27 &1.5$\times$1.3 & 56 &$\sim$2.0\\
\hline
\end{tabular}
\end{table*}

Below follows a brief description of the selected targets.\\
\textbf{N113}, located in the central part of the LMC, contains the most intense
H$_2$O maser of the Magellanic Clouds \citep{Ellingsen2010}, which indicates active
star formation activity at an early stage. A large number of molecular transitions have been reported in N113
(e.g., \citealt{Chin1996,Chin1997,Heikkila1998,Wang2009,Paron2014,Nishimura2016,Tang2017b,Sewilo2018,Sewilo2019}),
while interferometric data from HCN, HCO$^+$, HNC, CH$_3$OH, and 1.3\,mm continuum
\citep{Wong2006,Seale2012,Sewilo2018,Sewilo2019} reveal a filamentary structure elongated roughly North-South
with a length of $\sim$6\,pc, including a few dense clumps with radius $\sim$0.5\,pc.\\
\textbf{N159W}, located at the southwestern tip of 30 Doradus, one of the most intense star-forming regions in the LMC,
contains a large number of O- and B-type stars, embedded young stellar objects (YSOs),
and ultracompact H\,{\scriptsize II} regions (e.g., \citealt{Jones2005,Farina2009,Chen2010,Carlson2012}).
It shows the brightest single-dish CO\,($J$\,=\,1--0) peak in the LMC (e.g., \citealt{Johansson1994,Fukui1999,Wong2011})
and has been frequently observed with a large number of molecular transitions
(e.g., \citealt{Johansson1994,Chin1996,Chin1997,Heikkila1998,Heikkila1999,Paron2014,Pineda2008,Lee2016,Nishimura2016,Tang2017b}).
High resolution observations reveal a complex filamentary distribution of molecular gas \citep{Seale2012,Fukui2015}.
The gas temperature obtained from the NH$_3$(2,2)/(1,1) line ratio at $\sim$19$''$ resolution is $\sim$16\,K \citep{Ott2010},
which is only about half the temperatures derived from low resolution ($\sim$30$''$) para-H$_2$CO and the dust \citep{Tang2017b}.
After deriving kinetic temperatures from our high resolution H$_2$CO data, this difference will be discussed
in more detail in Sect.\,\ref{sect:Comparison-Tk}.

\subsection{Observations and data reduction}
\label{sect:Observations}
Our observations were carried out  with the 12\,m  array of
the Atacama Large Millimeter/submillimeter Array (ALMA) at Band\,6 (Project:\,2013.1.01136.S).
Two Execution Blocks were conducted using 35 and 38 antennas in January 17 and April 8, 2015, respectively.
The projected baseline lengths ranged from 15 to 349\,m. The primary beam (FWHM) and the maximum recoverable
scale are $\sim$$29''$ and $\sim$$13''$, respectively. The flux, bandpass, and phase calibrators were
J0519-454, J0519-4546, and J0635-7516, respectively. Three spectral windows were centered on
SiO\,(5--4), H$_2$CO\,(3$_{22}$--2$_{21}$), and CO\,(2--1), respectively, with a $\sim$937.5\,MHz bandwidth
and 960 channels for each band, yielding a channel width of 976.562\,kHz, corresponding to 1.27--1.35\,km\,s$^{-1}$.
For the para-H$_2$CO triplet the synthesized beam size is
$\sim$1\hbox{$\,.\!\!^{\prime\prime}$}6$\times$1\hbox{$\,.\!\!^{\prime\prime}$}3\,($\sim$0.4$\times$0.3\,pc$^2$ at 50\,kpc distance).
The surveyed areas of N113 and N159W are centered on $\alpha_{2000}$\,=\,05:13:18.2, $\delta_{2000}$\,=\,--69:22:25.0
and $\alpha_{2000}$\,=\,05:39:36.0, $\delta_{2000}$\,=--\,69:45:25.0,
respectively. Detailed observational parameters are listed in Table\,\ref{table:Observational_parameters}.
The para-H$_2$CO $J_{\rm K_aK_c}$\,=\,3$_{03}$--2$_{02}$, 3$_{22}$--2$_{21}$, and 3$_{21}$--2$_{20}$ transitions
have rest frequencies of 218.222, 218.475, and 218.760\,GHz, respectively.

\begin{figure*}[t]
\centering
\includegraphics[width=1.0\textwidth]{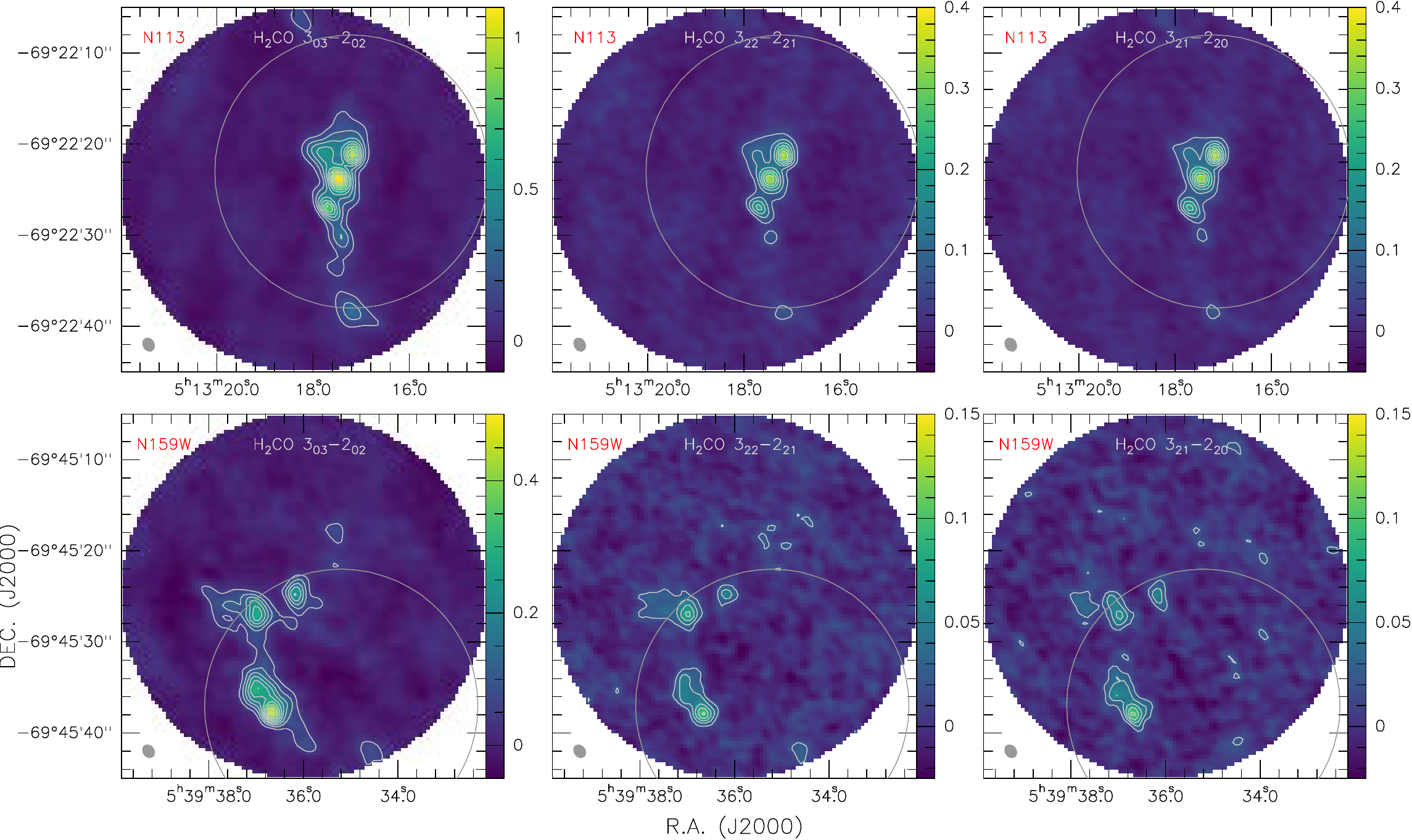}
\caption{Velocity-integrated intensity maps (color bars in units of
Jy\,beam$^{-1}$\,km\,s$^{-1}$) of para-H$_2$CO\,3$_{03}$--2$_{02}$ (\emph{left panels}),
3$_{22}$-2$_{21}$ (\emph{central panels}), and 3$_{21}$--2$_{20}$ (\emph{right panels}) of
N113 (Integrated Local Standard of Rest velocity range: 229--240\,km\,s$^{-1}$; \emph{top panels}) and N159W
(integrated velocity range: 232--244\,km\,s$^{-1}$; \emph{bottom panels}).
The centers of the fields are those also given in Sect.\,\ref{sect:Observations}, i.e.
$\alpha_{2000}$\,=\,05:13:18.2, $\delta_{2000}$\,=\,--69:22:25.0
and $\alpha_{2000}$\,=\,05:39:36.0, $\delta_{2000}$\,=--\,69:45:25.0.
The contours provide, with respect to the peak intensity, levels from 10\% to 100\% with steps
of 10\%\,($\sim$10$\sigma$ and $\sim$7$\sigma$) for para-H$_2$CO\,3$_{03}$--2$_{02}$ in N113 and N159W,
and from 10\% to 100\% with steps of 10\%\,($\sim$5$\sigma$) for para-H$_2$CO\,3$_{22}$-2$_{21}$ and 3$_{21}$--2$_{20}$.
The para-H$_2$CO\,3$_{03}$--2$_{02}$ peak intensities are 1.2 and 0.5\,Jy\,beam$^{-1}$\,km\,s$^{-1}$,
the para-H$_2$CO\,3$_{22}$-2$_{21}$ peak intensities are 0.5 and 0.2\,Jy\,beam$^{-1}$\,km\,s$^{-1}$,
and the para-H$_2$CO\,3$_{21}$--2$_{20}$ peak intensities are 0.5 and 0.2\,Jy\,beam$^{-1}$\,km\,s$^{-1}$
in N113 and N159W, respectively. The pixel size of each image is
0\hbox{$\,.\!\!^{\prime\prime}$}2$\times$0\hbox{$\,.\!\!^{\prime\prime}$}2.
The beam size of each image is shown in the lower left corner.
Gray circles show the half power beam sizes of the N113 and N159W APEX-12m data, reported by \cite{Tang2017b}.
The shown fields of view encompass slightly more than our ALMA primary half power
beam width (Sect.\,\ref{sect:Observations}). No primary beam correction has been applied.
Angular scales of 4$''$ correspond to a linear scale of $\sim$1\,pc.}
\label{fig:N113+N159W-maps}
\end{figure*}

The data were calibrated using the CASA\footnote{\tiny https://casa.nrao.edu} version 4.2.2
pipeline and were then imaged using CASA version 5.6.1. We imaged the calibrated data using
a modified script associated with the data release. No primary beam correction has been applied
for Figs.\,\ref{fig:N113+N159W-maps}, \ref{fig:N113-channel-maps}, and \ref{fig:N159W-channel-maps},
but it has been applied for data analysis. Data analysis associated with spectral lines and images was performed using
GILDAS\footnote{\tiny http://www.iram.fr/IRAMFR/GILDAS}. A typical rms noise level is
$\sim$2.0\,mJy\,beam$^{-1}$ for a 1.34\,km\,s$^{-1}$ wide channel (see Table\,\ref{table:Observational_parameters}).

As mentioned in Sect.\,\ref{sect:Targets}, the para-H$_2$CO\,(3--2) triplet has been observed before
with the APEX 12\,m telescope in N113 and N159W \citep{Tang2017b}. We integrate all the emission within
our ALMA data and compare it to the single-dish APEX 12\,m data (beam size $\sim$30$''$).
The fluxes recovered by our interferometer data are
$\sim$75\%, $\sim$80\%, and $\sim$80\% for para-H$_2$CO\,3$_{03}$--2$_{02}$, 3$_{21}$--2$_{20}$,
and 3$_{22}$--2$_{21}$, respectively, in N113. We find that $\sim$45\%, $\sim$50\%, and $\sim$50\% of
the para-H$_2$CO\,3$_{03}$--2$_{02}$, 3$_{21}$--2$_{20}$, and 3$_{22}$--2$_{21}$ integrated flux
observed by the APEX 12\,m telescope is recovered for N159W by our ALMA data, respectively.
These indicate that the para-H$_2$CO\,(3--2) triplets have, for a given source, similar
missing flux percentages.

We note that our observational field centers of N113 and N159W have $\sim$6" and $\sim$13"
offsets with respect to those used by \cite{Tang2017b}, respectively. This may affect the comparison
of our ALMA and single-dish data, especially for N159W. Our central positions are farther to the
east (N113) and north (N159W) of the bulk of the emission shown in Fig.\,\ref{fig:N113+N159W-maps},
so that part of the missing flux in our ALMA data may not be due to a lack of short baselines
but due to the chosen field centers. In the optically thin regime, the critical density of
the para-H$_2$CO\,3$_{03}$--2$_{02}$ transition is $\sim$6$\times$10$^5$\,cm$^{-3}$ at kinetic
temperature 50\,K \citep{Shirley2015}, so the para-H$_2$CO\,(3--2) triplet probes compact emission regions.
Considering that the para-H$_2$CO\,(3--2) triplet is a dense gas tracer and has similar missing
flux percentages within N113 and N159W, the line ratios of the para-H$_2$CO\,(3--2) transitions
are likely only weakly influenced by missing flux.

\section{Results}
\label{sect:Results}
\subsection{Overview}
The integrated intensity distributions of the para-H$_2$CO triplet in N113
and N159W are shown in Fig.\,\ref{fig:N113+N159W-maps}.
The observed spectra from the H$_2$CO peaks in N113 and N159W are shown in
Fig.\,\ref{fig:N113+N159W-Clump-spectrum}. The location of the seven dense H$_2$CO clumps
in each of the two massive star forming regions, selected by eye, are indicated in
Figs.\,\ref{fig:N113-Tk-H2CO-ratios} and \ref{fig:N159W-Tk-H2CO-ratios}, left panels,
and are listed in Table\,\ref{table:Clumps-Parameters}.

\subsection{Distribution of H$_2$CO}
\label{sect:h2CO-distribution}
In both sources, N113 and N159W, para-H$_2$CO\,(3$_{03}$--2$_{02}$) shows extended
distributions and clearly traces dense molecular structure,
which is consistent with previous observational results probed with
other gas tracers of e.g. $^{13}$CO, HCN, HNC, HCO$^+$
\citep{Wong2006,Seale2012,Fukui2015,Sewilo2018}.
The para-H$_2$CO 3$_{22}$--2$_{21}$ and 3$_{21}$--2$_{20}$ lines
are only detected in the densest regions of N113 and N159W,
and show less extended distributions than the para-H$_2$CO 3$_{03}$--2$_{02}$ transition.

The para-H$_2$CO 3$_{03}$--2$_{02}$ velocity channel maps of N113 and N159W are
shown in Figs.\,\ref{fig:N113-channel-maps} and \ref{fig:N159W-channel-maps} of Appendix\,A,
respectively. In N113, the filamentary structure, roughly elongated in
North-South direction with a length of $\sim$25$''$, is
prominent at low velocities, $\sim$232--236\,km\,s$^{-1}$,
while the northeastern clump (clump\,1 following the nomenclature of
Fig.\,\ref{fig:N113-Tk-H2CO-ratios} and Table\,\ref{table:Clumps-Parameters})
becomes visible at higher velocities, $\sim$236--238\,km\,s$^{-1}$.
In the N159W region, there appear to be two velocity components centered
at $\sim$236 and $\sim$240\,km\,s$^{-1}$ spreading from the northeast to the southwest.

\subsection{H$_2$CO line ratios}
\label{sect:H$_2$CO-line-ratio}
\begin{figure}[t]
\centering
\includegraphics[width=0.48\textwidth]{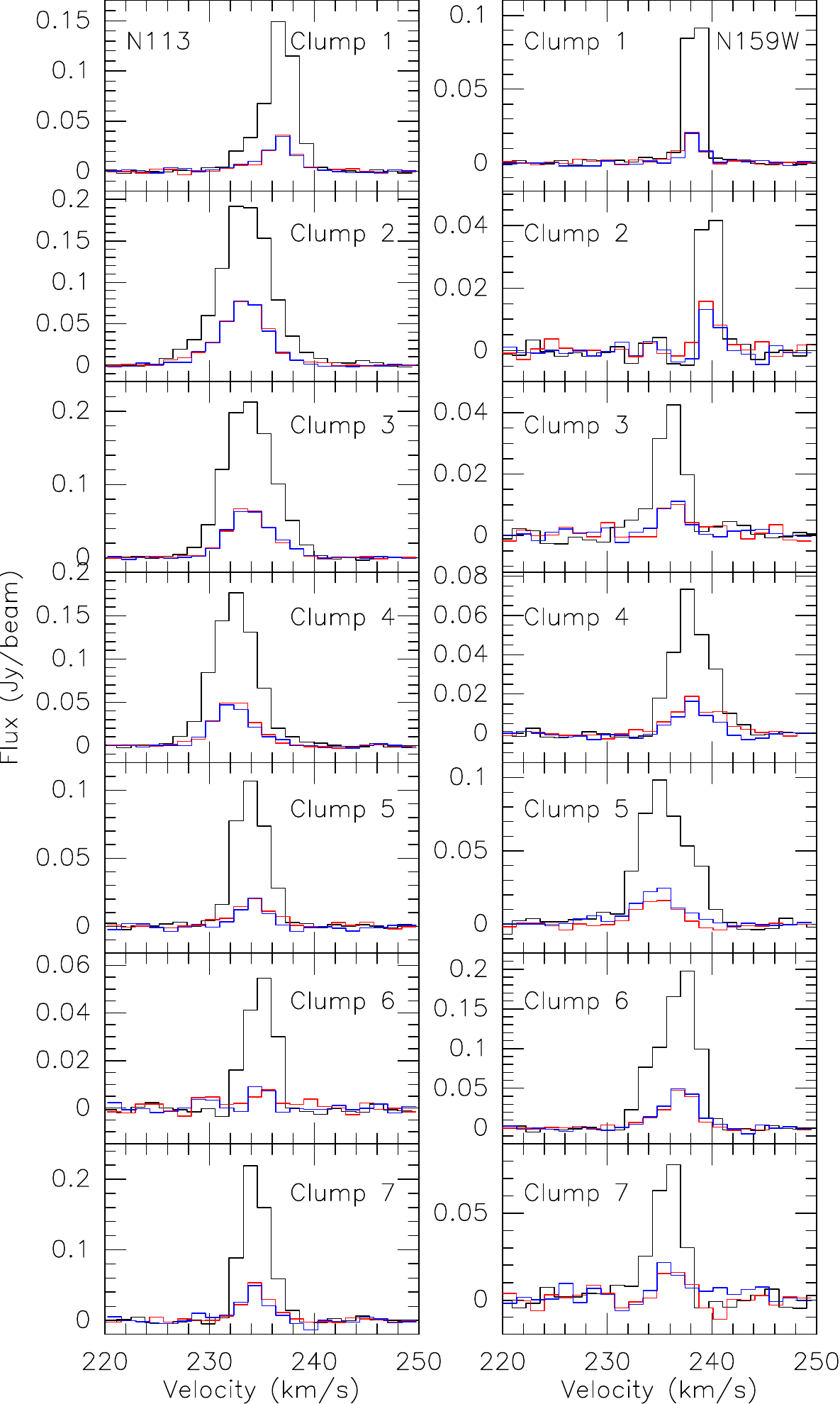}
\caption{Spectra towards H$_2$CO clumps in N113 (\emph{left panels}) and
N159W (\emph{right panels}; see Table\,\ref{table:Clumps-Parameters} and, for the numbering,
the left panels of Figs.\,\ref{fig:N113-Tk-H2CO-ratios} and \ref{fig:N159W-Tk-H2CO-ratios})
taken from the strongest pixel obtained with ALMA. The pixel size is
0\hbox{$\,.\!\!^{\prime\prime}$}2$\times$0\hbox{$\,.\!\!^{\prime\prime}$}2.
Black: para-H$_2$CO\,3$_{03}$--2$_{02}$, red: para-H$_2$CO\,3$_{22}$--2$_{21}$,
and blue: para-H$_2$CO\,3$_{21}$--2$_{20}$. The spectra are not smoothed, i.e.
the channel widths are those given in Table\,\ref{table:Observational_parameters}.}
\label{fig:N113+N159W-Clump-spectrum}
\end{figure}

As mentioned in Sect.\,\ref{sect:Introduction}, the intensity
ratio of para-H$_2$CO lines involving different $K_{\rm a}$ ladders yields estimates
of the kinetic temperature of the gas \citep{Mangum1993a}. Generally, higher line ratios of
para-H$_2$CO\,(3$_{22}$--2$_{21}$/3$_{03}$--2$_{02}$ and 3$_{21}$--2$_{20}$/3$_{03}$--2$_{02}$)
indicate higher kinetic temperatures (e.g., \citealt{Mangum1993a,Ao2013,Ginsburg2016,Tang2017b,Tang2018a,Tang2018b};
see also Fig.\,\ref{fig:Tk-H2CO-ratios} for a more detailed view).
Therefore, the ratio maps can be used as a proxy for relative kinetic temperature.
The upper levels of the para-H$_2$CO\,3$_{22}$--2$_{21}$ and 3$_{21}$--2$_{20}$
transitions have almost the same energy above the ground state ($E_{\rm u}$\,$\simeq$\,68\,K).
They show similar spatial distributions (see Fig.\,\ref{fig:N113+N159W-maps}),
similar line profiles (for brightness temperature, linewidth,
and velocity in our observations; see Fig.\,\ref{fig:N113+N159W-Clump-spectrum}
and also \citealt{Tang2017a,Tang2017b,Tang2018a,Tang2018b}),
and are often detected simultaneously in molecular clouds
(e.g., \citealt{Bergman2011,Wang2012,Lindberg2012,Ao2013,Immer2014,Trevino2014,Ginsburg2016,Ginsburg2017,Tang2017a,Tang2017b,Tang2018a,Tang2018b,Lu2017,Lu2021,Mangum2019}).
Therefore, we use the averaged ratio
para-H$_2$CO\,0.5$\times$[(3$_{22}$--2$_{21}$\,+\,3$_{21}$--2$_{20})$/3$_{03}$--2$_{02}$]
between para-H$_2$CO\,3$_{22}$--2$_{21}$/3$_{03}$--2$_{02}$ and
3$_{21}$--2$_{20}$/3$_{03}$--2$_{02}$ for the first highly qualitative analysis below.

Averaged para-H$_2$CO\,(3--2) line ratio maps of N113 and N159W
are shown in the left panels of Figs.\,\ref{fig:N113-Tk-H2CO-ratios}
and \ref{fig:N159W-Tk-H2CO-ratios}, respectively.
The line ratios are calculated by velocity-integrated intensities where
the para-H$_2$CO\,3$_{22}$--2$_{21}$ and 3$_{21}$--2$_{20}$ lines are
detected with signal-to-noise ratios (S/N) $\gtrsim$\,3$\sigma$. Para-H$_2$CO line ratios range from
0.10 to 0.38 with an average of 0.22\,$\pm$\,0.01 (errors given here and elsewhere
are standard deviations of the mean) in N113 and from 0.10 to 0.28
with an average of 0.20\,$\pm$\,0.01 in N159W (see also Table\,\ref{table:Parameters}).
The lower ratios ($<$0.2) occur in the outskirts of the N113 and N159W H$_2$CO clouds
(see Figs.\,\ref{fig:N113-Tk-H2CO-ratios} and \ref{fig:N159W-Tk-H2CO-ratios}).
Higher ratios ($\gtrsim$0.2) associate with dense clumps, YSOs (or YSO
candidates) \citep{Chen2010,Carlson2012}, and/or H$_2$O masers \citep{Ellingsen2010}
in N113 and N159W.

\subsection{Kinetic temperatures from H$_2$CO line ratios}
\label{sect:Kinetic-temperature}
According to Sect.\,\ref{sect:H$_2$CO-line-ratio}, the para-H$_2$CO\,3$_{22}$--2$_{21}$
and 3$_{21}$--2$_{20}$ transitions have almost the same upper-state energies above the ground state
and similar observed line profiles. Para-H$_2$CO\,3$_{22}$--2$_{21}$/3$_{03}$--2$_{02}$ and
3$_{21}$--2$_{20}$/3$_{03}$--2$_{02}$ ratios are both good thermometers
to determine kinetic temperatures ($T_{\rm kin}$\,<\,150\,K; \citealt{Mangum1993a,Tang2018b})
and show a similar behavior to kinetic temperature and spatial density at densities
$n$(H$_2$)\,$\gtrsim$\,10$^5$\,cm$^{-3}$ \citep{Lindberg2015,Tang2017a}.
Therefore, we continue to use the averaged para-H$_2$CO ratio
0.5$\times$[(3$_{22}$--2$_{21}$\,+\,3$_{21}$--2$_{20})$/3$_{03}$--2$_{02}$]
to directly determine gas kinetic temperatures.

\begin{figure}[t]
\vspace*{0.2mm}
\begin{center}
\includegraphics[width=0.48\textwidth]{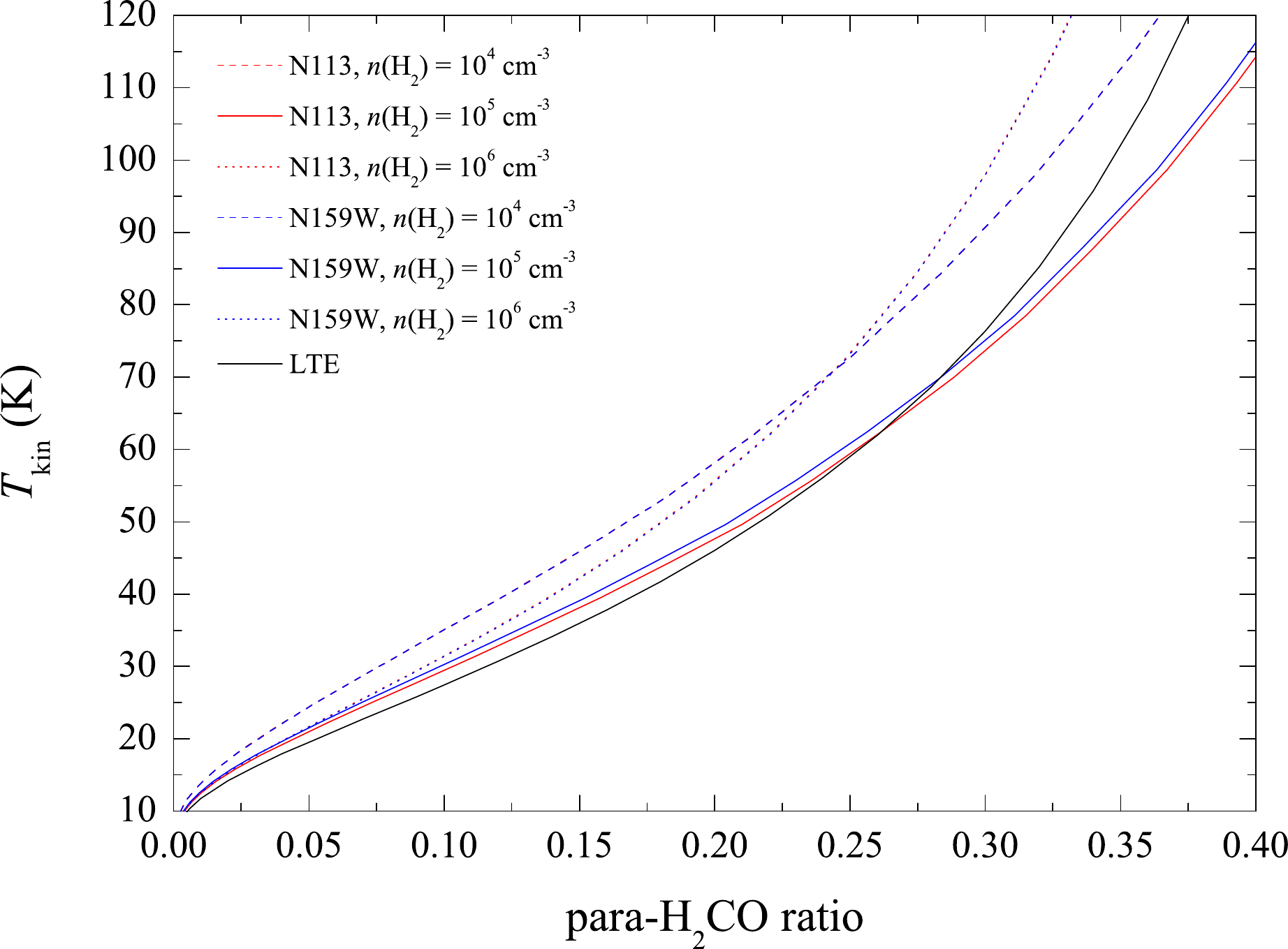}
\end{center}
\caption{RADEX non-LTE (\emph{red and blue lines}) and LTE (\emph{black line}) modeling of the relation between
kinetic temperature and average ratio of para-H$_2$CO\,3$_{22}$--2$_{21}$/3$_{03}$--2$_{02}$ and
3$_{21}$--2$_{20}$/3$_{03}$--2$_{02}$ with spatial densities of
$n$(H$_2$)\,=\,10$^{4}$, 10$^{5}$, and 10$^{6}$\,cm$^{-3}$ (\emph{dashed, solid, and dotted lines, respectively}),
an averaged observed linewidth of 4.0\,km\,s$^{-1}$, and column densities of $N$(para-H$_2$CO)\,=\,2.7$\times$10$^{12}$
and 3.7$\times$10$^{12}$\,cm$^{-2}$ for N113 (\emph{red lines}) and N159W (\emph{blue lines}) \citep{Tang2017b}, respectively.
$T_{\rm kin}$-Ratio(H$_2$CO) relations for N113 and N159W at density 10$^{4}$\,cm$^{-3}$ are nearly the same,
which is similar to the situation at 10$^{6}$\,cm$^{-3}$.}
\label{fig:Tk-H2CO-ratios}
\end{figure}

Using the RADEX\footnote{\tiny http://var.sron.nl/radex/radex.php}
non-LTE modeling program \citep{van2007} with collisional rate coefficients from \cite{Wiesenfeld2013},
we modeled the relation between the gas kinetic temperature and the measured average of
para-H$_2$CO\,0.5$\times$[(3$_{22}$--2$_{21}$\,+\,3$_{21}$--2$_{20})$/3$_{03}$--2$_{02}$]
ratios, adopting a 2.73\,K background temperature, an average observational linewidth of 4.0\,km\,s$^{-1}$,
and column densities $N$(para-H$_2$CO)\,=\,2.7$\times$10$^{12}$ and 3.7$\times$10$^{12}$\,cm$^{-2}$
for N113 and N159W, respectively. The results are shown in Fig.\,\ref{fig:Tk-H2CO-ratios}.
The values of the para-H$_2$CO column density were obtained with APEX data (beam size $\sim$ 30$''$; \citealt{Tang2017b}),
which cover similar regions. Different column densities of para-H$_2$CO only weakly
affect derived kinetic temperatures (see Fig.\,3 in \citealt{Tang2017b} or Fig.\,4 in \citealt{Tang2018a};
this was also shown in Fig.\,13 and discussed in Sect.\,4.3.1 of \citealt{Mangum1993a})
as long as all lines are optically thin. Considering that the relation between the gas temperature
and the para-H$_2$CO line ratio may vary at different spatial densities (see Fig.\,2 in \citealt{Tang2017b}),
we modeled it at spatial densities 10$^{4}$, 10$^{5}$, and 10$^{6}$\,cm$^{-3}$ in Fig.\,\ref{fig:Tk-H2CO-ratios}.
It appears that $T_{\rm kin}$ at $n$(H$_2$)\,=\,10$^{5}$\,cm$^{-3}$ is consistently lower than values at
10$^{4}$ and 10$^{6}$\,cm$^{-3}$ by $\lesssim$23\% and $\lesssim$34\%, respectively, for $T_{\rm kin}$\,$\lesssim$\,100\,K.
Local thermodynamic equilibrium (LTE) is a good approximation for the H$_2$CO level populations under optically thin
and high-density conditions \citep{Mangum1993a,Tang2017a,Tang2017b,Tang2018b}. Following the method applied by \cite{Tang2017b}
in their Eq.\,(2), we plot the relation between the LTE kinetic temperature, $T_{\rm LTE}$, and the para-H$_2$CO\,(3--2) line ratio
in Fig.\,\ref{fig:Tk-H2CO-ratios}. Apparently, $T_{\rm LTE}$ agrees well with $T_{\rm non-LTE}$ at volume densities
$n$(H$_2$)\,$\sim$\,10$^{5}$\,cm$^{-3}$ as long as $T_{\rm kin}$\,$\lesssim$\,100\,K.
Previous observations show that para-H$_2$CO\,(3--2) is sensitive to gas temperature
at density 10$^{5}$\,cm$^{-3}$ \citep{Ginsburg2016,Immer2016,Tang2017b}.
The spatial density measured with para-H$_2$CO\,(3$_{03}$--2$_{02}$) and C$^{18}$O\,(2--1) in N113
and N159W is $n$(H$_2$)\,$\sim$\,10$^{5}$\,cm$^{-3}$ on a size of $\sim$30$''$ \citep{Tang2017b}.
Therefore, here we adopt 10$^{5}$\,cm$^{-3}$ as an averaged spatial gas
density in the N113 and N159W regions.

\begin{figure*}[t]
\vspace*{0.2mm}
\begin{center}
\includegraphics[width=0.98\textwidth]{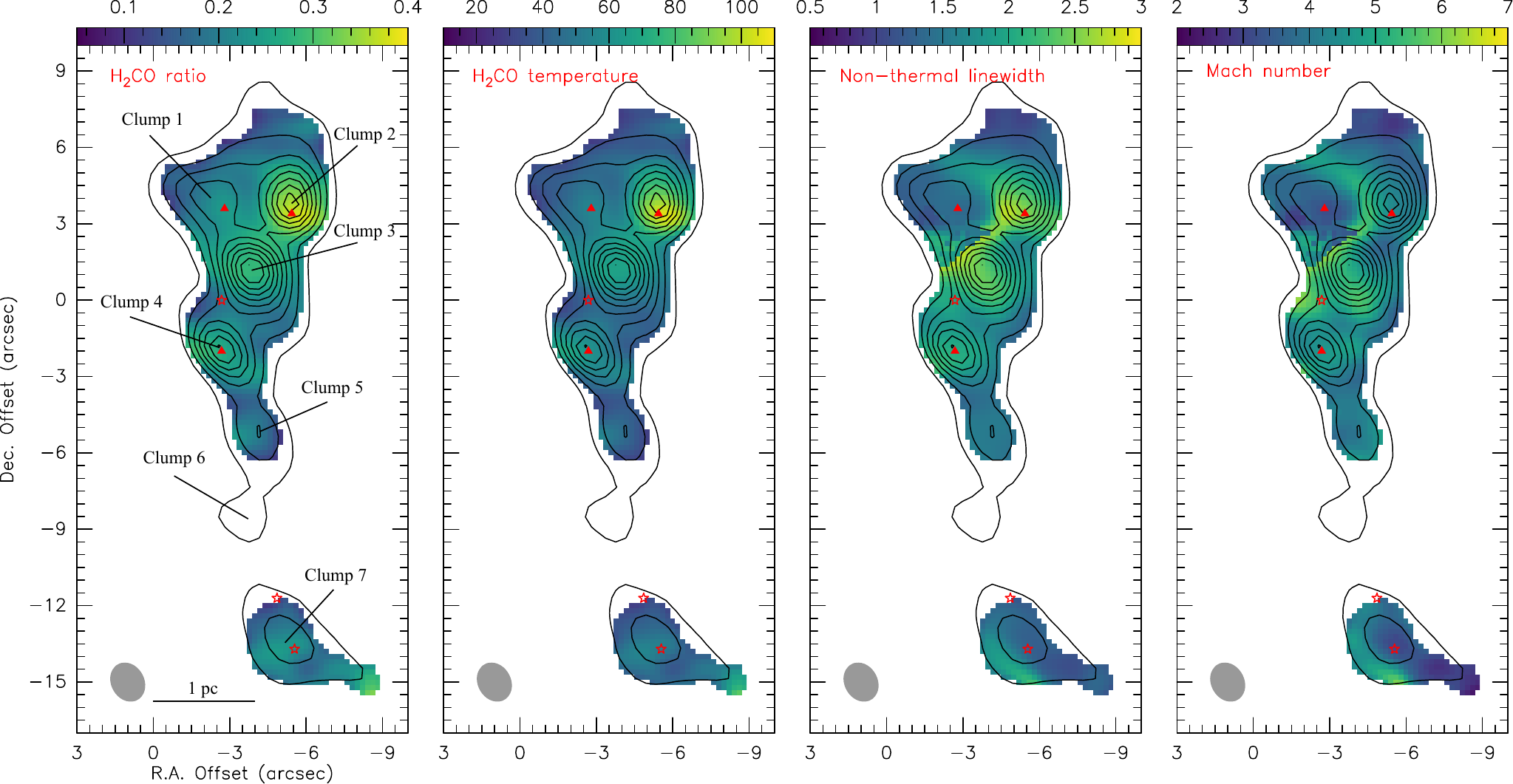}
\end{center}
\caption{\emph{Left two panels}: The averaged velocity-integrated intensity ratio map of
para-H$_2$CO\,0.5$\times$[(3$_{22}$--2$_{21}$\,+\,3$_{21}$--2$_{20})$/3$_{03}$--2$_{02}$]
in N113. The kinetic temperatures derived from the para-H$_2$CO\,(3--2) line ratios (color bar in units of Kelvin).
\emph{Right two panels}: Maps of non-thermal velocity dispersion (color bar in units of km\,s$^{-1}$)
and Mach number. Black contours show levels of integrated intensity of the
para-H$_2$CO\,(3$_{03}$--2$_{02}$) line (see Fig.\,\ref{fig:N113+N159W-maps}).
Offsets are relative to our reference position for N113 (see Fig.\,\ref{fig:N113+N159W-maps}).
The pixel size of each image is 0\hbox{$\,.\!\!^{\prime\prime}$}2$\times$0\hbox{$\,.\!\!^{\prime\prime}$}2.
The beam size of each image is shown in the lower left corner. Stars and triangles show the locations of
YSOs (or YSO candidates) and H$_2$O masers \citep{Chen2010,Ellingsen2010,Carlson2012}, respectively.}
\label{fig:N113-Tk-H2CO-ratios}
\end{figure*}

\begin{figure*}[t]
\vspace*{0.2mm}
\begin{center}
\includegraphics[width=0.982\textwidth]{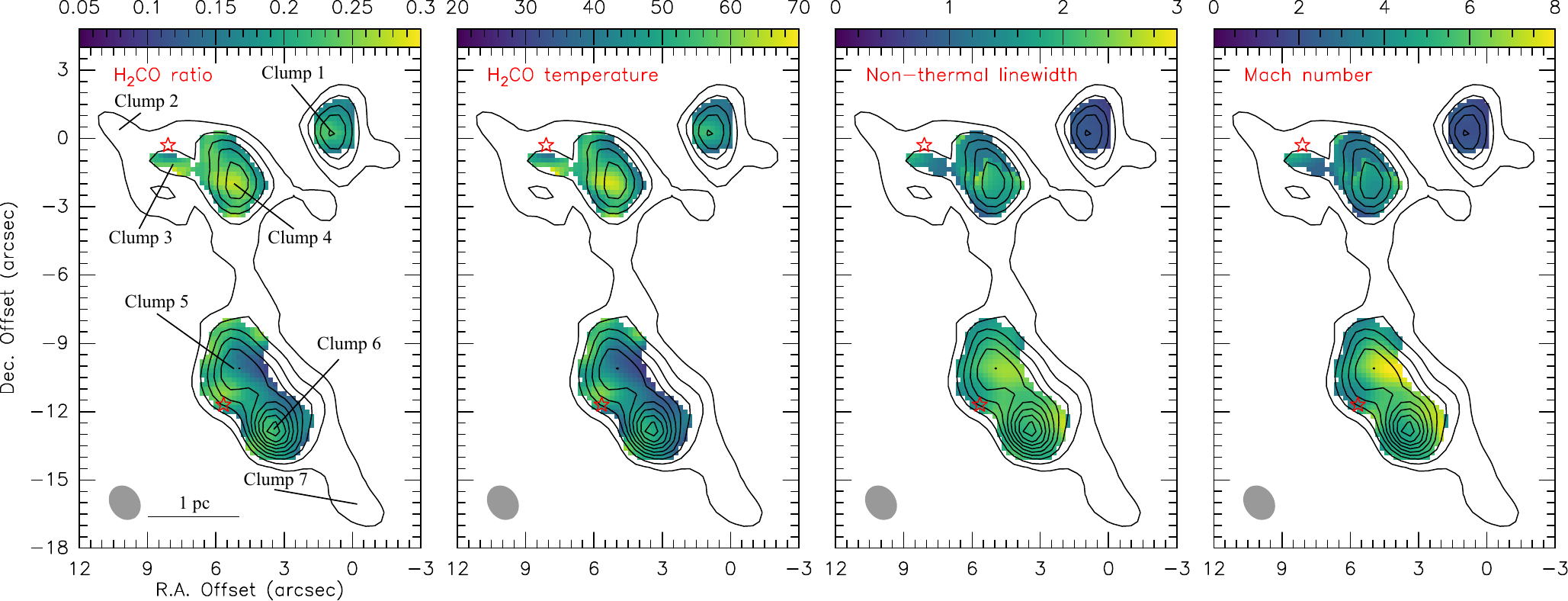}
\end{center}
\caption{Same as Fig.\,\ref{fig:N113-Tk-H2CO-ratios}, but for N159W.}
\label{fig:N159W-Tk-H2CO-ratios}
\end{figure*}

We use the relation between kinetic temperature and para-H$_2$CO line ratios at averaged
spatial density 10$^5$\,cm$^{-3}$and averaged column densities of N113
and N159W (shown in Fig.\,\ref{fig:Tk-H2CO-ratios}) to convert ratio maps to temperature
maps in Figs.\,\ref{fig:N113-Tk-H2CO-ratios} and \ref{fig:N159W-Tk-H2CO-ratios}.
The gas kinetic temperatures derived from the para-H$_2$CO line ratios are warm at density
$n$(H$_2$)\,=\,10$^5$\,cm$^{-3}$, ranging from 28 to 105\,K in N113 and from 29 to 68\,K in N159W on a scale
of $\sim$0.4\,pc (see also Table\,\ref{table:Parameters}).
It seems that higher temperatures ($T_{\rm kin}$\,$\gtrsim$\,50\,K) associate with dense clumps,
YSOs (or YSO candidates), and/or H$_2$O masers in N113 and N159W
(see Figs.\,\ref{fig:N113-Tk-H2CO-ratios} and \ref{fig:N159W-Tk-H2CO-ratios}).
Gas with lower temperatures ($T_{\rm kin}$\,$<$\,50\,K) is located at the outskirts of the H$_2$CO
distributions of both N113 and N159W. The kinetic temperatures of the dense gas are high ($T_{\rm kin}$\,=\,44--96\,K)
in the dense H$_2$CO clumps (see Table\,\ref{table:Clumps-Parameters}), which is consistent with previous
observational results measured with para-H$_2$CO\,($J$=3--2 and 4--3) from dense clumps representing
various evolutionary stages of high-mass star formation in the Galactic plane \citep{Tang2018b}.
The gas kinetic temperatures show apparent gradients in these dense H$_2$CO clumps of N113 and N159W
(e.g., clumps 2, 3, 4, 5, and 7 in N113 and clumps 1, 4, and 6 in N159W; see Figs.\,\ref{fig:N113-Tk-H2CO-ratios}
and \ref{fig:N159W-Tk-H2CO-ratios}) on a clump scale of $\sim$1\,pc.

As mentioned in Sect.\,\ref{sect:Targets}, the dense gas kinetic temperatures in N113 and N159W
have been estimated with the same transitions of para-H$_2$CO\,($J$=3--2) observed by the APEX 12-m telescope
(beam size $\sim$30$''$; \citealt{Tang2017b}). For a comparison with the lower resolution APEX data,
we have therefore averaged our data for areas, where the presence of the dense gas shown in
Figs.\,\ref{fig:N113-Tk-H2CO-ratios} and \ref{fig:N159W-Tk-H2CO-ratios} could be identified
by the para-H$_2$CO\,3$_{03}$--2$_{02}$ line. The averaged temperatures are 48 and 36\,K in N113 and N159W,
respectively, which are consistent with previous results in N113 (54\,$\pm$\,7\,K) and
N159W (35\,$\pm$\,4\,K) obtained with the single-dish observations \citep{Tang2017b}.

Previous observations toward the Galactic central molecular zone (CMZ) clouds and dense
massive star-forming clumps in the Galactic plane show that the opacities of
para-H$_2$CO\,($J$=3--2) lines influence only weakly the measurements of gas kinetic temperature
\citep{Ginsburg2016,Immer2016,Tang2018b}. Nevertheless, a major caveat in our discussion are
the poorly known opacities of para-H$_2$CO\,($J$=3--2) lines in N113 and N159W. While the
para-H$_2$CO\,3$_{22}$--2$_{21}$ and 3$_{21}$--2$_{20}$ lines,
being much weaker than the 3$_{03}$--2$_{02}$ line, appear to be optically thin,
this is not necessarily the case for our strongest emitter, the 3$_{03}$--2$_{02}$ transition.
If this line were optically thick in the molecular cores we have observed,
line ratios would be too close to unity, and we would overestimate, relative to
the optically thin case, the para-H$_2$CO\,0.5$\times$[(3$_{22}$--2$_{21}$\,+\,3$_{21}$--2$_{20})$/3$_{03}$--2$_{02}$]
intensity ratios and kinetic temperatures. This implies that in case of saturation we would obtain
upper limits to the kinetic temperature only. This could easily mimic the high temperatures in the cores
and the lower temperatures in the optically thin outskirts of the N113 and N159W clouds, we have
mentioned above, possibly providing a good agreement with the low kinetic temperature
derived from ammonia (beam size 18\hbox{$\,.\!\!^{\prime\prime}$}6$\times$15\hbox{$\,.\!\!^{\prime\prime}$}7;
\citealt{Ott2010}), since the gas traced by ammonia presumably resides in a well shielded highly obscured cloud core.

A first estimate of optical depth can be attained by a comparison of para-H$_2$CO\,3$_{03}$--2$_{02}$
main beam brightness temperatures with the temperatures derived from the dust and from the line
temperature ratios. On our 0.4\,pc scale, observed main beam brightness temperatures should be close to
the real brightness temperatures of the emission (e.g. \citealt{Tang2018a}), which should reach values
of order of 16\,K (the $T_{\rm kin}$ value from the NH$_3$(2,2)/(1,1) line ratio; \citealt{Ott2010})
or higher to reveal strong indications of saturation. Converting with
$T_{\rm mb}$\,=\,$S$\,$\times$\,$\lambda$$^2$/(2.65\,$\times$\,$\theta_b$$^2$)
($T_{\rm mb}$: main beam brightness temperature in units of K; $S$: flux density in units of Jy; $\lambda$:
wavelength in units of cm; $\theta_b$: our beam size in units of arcmin) the flux density scale into
units of main beam brightness temperature and accounting for typical total linewidths of
$\sim$4.0\,km\,s$^{-1}$ in case of N113 and N159W (see Table\,\ref{table:Clumps-Parameters}),
we obtain peak temperatures of only $\sim$2.5\,K. This is well below the low ammonia temperature from \cite{Ott2010}
and indicates either highly subthermal excitation with potentially saturated para-H$_2$CO\,3$_{03}$--2$_{02}$ line
emission or a purely optically thin scenario.

\begin{table*}[t]
\scriptsize
\caption{Parameters of H$_2$CO clumps derived from N113 and N159W.}
\centering
\begin{tabular}
{ccccccccccccccccc}
\hline\hline
Source & Clump & Offset &Transition &$\int$Flux\,d$v$ &$V_{\rm LSR}$ &FWHM &Flux & $\sigma_{\rm T}$ & $\sigma_{\rm NT}$ & $c_{\rm s}$  & $\mathcal{M}$ & $T_{\rm kin}$ & $T_{\rm turb}$ \\
       & & " , " &  &Jy\,km\,s$^{-1}$ &km\,s$^{-1}$ &km\,s$^{-1}$ &Jy &km\,s$^{-1}$ & km\,s$^{-1}$ & km\,s$^{-1}$ & & K & K \\
\hline
N113   &1 & --2.4, 4.0    &3$_{03}$--2$_{02}$ &0.530 (0.006) &236.9 (0.1) &3.3 (0.1) &0.150 & 0.11 (0.01) & 1.41 (0.02) & 0.40 (0.02) & 3.5 & 50.4 (5.1) & 50 \\
       &  &               &3$_{22}$--2$_{21}$ &0.110 (0.006) &236.9 (0.1) &3.0 (0.2) &0.035 &             &             &             &     &            &    \\
       &  &               &3$_{21}$--2$_{20}$ &0.114 (0.007) &236.7 (0.1) &3.4 (0.3) &0.032 &             &             &             &     &            &    \\
       &2 & --5.4, 3.8    &3$_{03}$--2$_{02}$ &1.226 (0.019) &233.5 (0.1) &5.8 (0.1) &0.198 & 0.16 (0.01) & 2.46 (0.04) & 0.56 (0.02) & 4.4 & 96.2 (5.8) & 93 \\
       &  &               &3$_{22}$--2$_{21}$ &0.429 (0.006) &233.3 (0.1) &5.2 (0.1) &0.077 &             &             &             &     &            &    \\
       &  &               &3$_{21}$--2$_{20}$ &0.435 (0.008) &233.4 (0.1) &5.2 (0.1) &0.078 &             &             &             &     &            &    \\
       &3 & --4.0, 1.2    &3$_{03}$--2$_{02}$ &1.258 (0.005) &233.7 (0.1) &5.5 (0.1) &0.216 & 0.13 (0.01) & 2.32 (0.01) & 0.47 (0.01) & 5.0 & 67.7 (2.6) & 67 \\
       &  &               &3$_{22}$--2$_{21}$ &0.350 (0.008) &233.8 (0.1) &4.8 (0.1) &0.068 &             &             &             &     &            &    \\
       &  &               &3$_{21}$--2$_{20}$ &0.345 (0.007) &233.9 (0.1) &5.0 (0.1) &0.065 &             &             &             &     &            &    \\
       &4 & --2.8, --2.0  &3$_{03}$--2$_{02}$ &0.886 (0.007) &232.4 (0.1) &4.8 (0.1) &0.173 & 0.12 (0.01) & 2.04 (0.02) & 0.44 (0.01) & 4.6 & 60.7 (2.3) & 60 \\
       &  &               &3$_{22}$--2$_{21}$ &0.231 (0.006) &232.3 (0.1) &4.2 (0.1) &0.051 &             &             &             &     &            &    \\
       &  &               &3$_{21}$--2$_{20}$ &0.225 (0.005) &232.2 (0.1) &4.4 (0.1) &0.048 &             &             &             &     &            &    \\
       &5 & --4.2, --5.2  &3$_{03}$--2$_{02}$ &0.409 (0.005) &233.9 (0.1) &3.6 (0.1) &0.108 & 0.11 (0.01) & 1.52 (0.02) & 0.38 (0.01) & 4.0 & 44.0 (2.5) & 44 \\
       &  &               &3$_{22}$--2$_{21}$ &0.090 (0.006) &234.2 (0.2) &4.6 (0.4) &0.019 &             &             &             &     &            &    \\
       &  &               &3$_{21}$--2$_{20}$ &0.073 (0.006) &234.2 (0.1) &3.2 (0.3) &0.022 &             &             &             &     &            &    \\
       &6 & --3.8, --8.6  &3$_{03}$--2$_{02}$ &0.203 (0.005) &234.9 (0.1) &3.3 (0.1) &0.057 & ...         &...          &...          &...  &...         &... \\
       &  &               &3$_{22}$--2$_{21}$ &...           &            &          &      &             &             &             &     &            &    \\
       &  &               &3$_{21}$--2$_{20}$ &...           &            &          &      &             &             &             &     &            &    \\
       &7 & --5.4, --13.4 &3$_{03}$--2$_{02}$ &0.710 (0.010) &234.3 (0.1) &3.0 (0.1) &0.220 & 0.12 (0.01) & 1.28 (0.02) & 0.41 (0.02) & 3.1 & 52.5 (4.5) & 52 \\
       &  &               &3$_{22}$--2$_{21}$ &0.159 (0.012) &234.5 (0.1) &2.9 (0.3) &0.052 &             &             &             &     &            &    \\
       &  &               &3$_{21}$--2$_{20}$ &0.165 (0.017) &234.3 (0.2) &3.2 (0.4) &0.049 &             &             &             &     &            &    \\
\hline
N159W  &1 & 0.8, 0.4      &3$_{03}$--2$_{02}$ &0.245 (0.003) &238.4 (0.1) &1.8 (0.1) &0.128 & 0.11 (0.01) & 0.79 (0.03) & 0.41 (0.01) & 1.9 & 51.8 (4.4) & 51 \\
       &  &               &3$_{22}$--2$_{21}$ &0.048 (0.003) &238.0 (0.1) &2.2 (0.1) &0.020 &             &             &             &     &            &    \\
       &  &               &3$_{21}$--2$_{20}$ &0.045 (0.003) &238.3 (0.1) &2.0 (0.2) &0.021 &             &             &             &     &            &    \\
       &2 & 10.4, 0.4.    &3$_{03}$--2$_{02}$ &0.114 (0.005) &239.7 (0.1) &1.6 (0.4) &0.066 & ...         &...          &...          &...  &...         &... \\
       &  &               &3$_{22}$--2$_{21}$ &...           &            &          &      &             &             &             &     &            &    \\
       &  &               &3$_{21}$--2$_{20}$ &...           &            &          &      &             &             &             &     &            &    \\
       &3 & 7.8, --1.2    &3$_{03}$--2$_{02}$ &0.141 (0.006) &236.0 (0.1) &3.1 (0.2) &0.043 & 0.12 (0.01) & 1.31 (0.07) & 0.43 (0.03) & 3.1 & 56.0 (9.2) & 55 \\
       &  &               &3$_{22}$--2$_{21}$ &0.030 (0.006) &236.4 (0.3) &2.6 (0.6) &0.011 &             &             &             &     &            &    \\
       &  &               &3$_{21}$--2$_{20}$ &0.035 (0.004) &236.3 (0.2) &2.8 (0.4) &0.012 &             &             &             &     &            &    \\
       &4 & 5.0, --2.0    &3$_{03}$--2$_{02}$ &0.309 (0.006) &238.0 (0.1) &4.3 (0.1) &0.068 & 0.13 (0.01) & 1.86 (0.05) & 0.46 (0.01) & 4.0 & 65.5 (5.1) & 65 \\
       &  &               &3$_{22}$--2$_{21}$ &0.104 (0.005) &238.5 (0.1) &5.9 (0.4) &0.017 &             &             &             &     &            &    \\
       &  &               &3$_{21}$--2$_{20}$ &0.081 (0.006) &238.3 (0.2) &4.5 (0.4) &0.017 &             &             &             &     &            &    \\
       &5 & 5.2, --10.4   &3$_{03}$--2$_{02}$ &0.527 (0.013) &235.4 (0.1) &5.4 (0.2) &0.091 & 0.11 (0.01) & 2.32 (0.08) & 0.38 (0.01) & 6.1 & 45.2 (4.0) & 46 \\
       &  &               &3$_{22}$--2$_{21}$ &0.101 (0.008) &234.8 (0.2) &4.8 (0.4) &0.020 &             &             &             &     &            &    \\
       &  &               &3$_{21}$--2$_{20}$ &0.115 (0.008) &234.8 (0.2) &4.6 (0.4) &0.023 &             &             &             &     &            &    \\
       &6 & 3.4, --13.0   &3$_{03}$--2$_{02}$ &0.914 (0.009) &236.9 (0.1) &4.5 (0.1) &0.190 & 0.11 (0.01) & 1.94 (0.03) & 0.41 (0.01) & 4.8 & 51.3 (3.3) & 51 \\
       &  &               &3$_{22}$--2$_{21}$ &0.189 (0.008) &237.0 (0.1) &3.6 (0.2) &0.049 &             &             &             &     &            &    \\
       &  &               &3$_{21}$--2$_{20}$ &0.200 (0.011) &237.0 (0.1) &3.8 (0.3) &0.049 &             &             &             &     &            &    \\
       &7 & --0.6, --16.4 &3$_{03}$--2$_{02}$ &0.272 (0.013) &235.8 (0.1) &3.1 (0.2) &0.081 & ...         &...          &...          &...  &...         &... \\
       &  &               &3$_{22}$--2$_{21}$ &...           &            &          &      &             &             &             &     &            &    \\
       &  &               &3$_{21}$--2$_{20}$ &...           &            &          &      &             &             &             &     &            &    \\
\hline
\end{tabular}
\label{table:Clumps-Parameters}
\tablefoot{Offsets relative to our reference positions for N113 and N159W (see Figs.\,\ref{fig:N113+N159W-maps},
\ref{fig:N113-Tk-H2CO-ratios}, and \ref{fig:N159W-Tk-H2CO-ratios}).
Velocity-integrated flux, $\int$Flux\,d$v$, local standard of rest velocity, $V_{\rm LSR}$, full width
at half maximum line width (FWHM), and peak flux (Flux) were obtained from Gaussian
fits using CLASS as part of the GILDAS software. For the thermal and non-thermal velocity dispersions,
$\sigma_{\rm T}$ and $\sigma_{\rm NT}$, the sound velocity $c_{\rm s}$, the Mach number $\mathcal{M}$,
the kinetic temperature $T_{\rm kin}$, and the turbulent temperature $T_{\rm turb}$,
see Sects.\,\ref{Sec:Non-thermal motions} and \ref{Sec:Turbulence}. Values in parenthesis are uncertainties.}
\end{table*}

\begin{table*}[t]
\caption{Thermal and non-thermal parameters derived from N113 and N159W.}
\centering
\begin{tabular}
{lccccccc}
\hline\hline
Parameter & \multicolumn{3}{c}{N113} & & \multicolumn{3}{c}{N159W}\\
\cline{2-4} \cline{6-8}
& Range & Median & Mean & & Range & Median & Mean\\
\hline
H$_2$CO line ratio                 & 0.10--0.38  & 0.21 & 0.22 $\pm$ 0.01 & & 0.10--0.28 & 0.20 & 0.20 $\pm$ 0.01 \\
$T_{\rm gas}$ /K                   & 27.6--105.4 & 48.5 & 51.4 $\pm$ 0.4  & & 28.7--67.8 & 47.5 & 47.6 $\pm$ 0.3  \\
$\sigma_{\rm T}$ /\,km\,s$^{-1}$   & 0.08--0.16  & 0.11 & 0.11 $\pm$ 0.01 & & 0.09--0.13 & 0.11 & 0.11 $\pm$ 0.01 \\
$\sigma_{\rm NT}$ /\,km\,s$^{-1}$  & 0.93--2.76  & 1.64 & 1.69 $\pm$ 0.01 & & 0.57--2.74 & 1.85 & 1.74 $\pm$ 0.02 \\
$c_{\rm s}$ /\,km\,s$^{-1}$        & 0.30--0.58  & 0.40 & 0.40 $\pm$ 0.01 & & 0.31--0.47 & 0.39 & 0.39 $\pm$ 0.01 \\
$\mathcal{M}$                      & 2.3--6.2    & 4.2  & 4.2 $\pm$ 0.1   & & 1.6--8.3   & 4.5  & 4.5  $\pm$ 0.1  \\
\hline 
\end{tabular}
\label{table:Parameters}
\tablefoot{For the meaning of the parameters in column 1, see Table\,\ref{table:Clumps-Parameters}. The errors shown
in columns 4 and 7 are the standard deviations of the mean.}
\end{table*}

To discriminate between these two possibilities, we used again the RADEX non-LTE program \citep{van2007}.
Setting $T_{\rm kin}$ to 16\,K, the low value derived by \cite{Ott2010} from ammonia, and a linewidth of
3\,\,km\,s$^{-1}$ we can reproduce the $T_{\rm mb}$\,(H$_2$CO\,3$_{03}$--2$_{02}$) line intensity of $\sim$2.5\,K with
$N$(para-H$_2$CO)\,=\,1.4\,$\times$\,10$^{14}$\,cm$^{-2}$, but the resulting
para-H$_2$CO\,0.5\,$\times$\,[(3$_{22}$--2$_{21}$\,+\,3$_{21}$--2$_{20}$)/3$_{03}$--2$_{02}$]
line ratio is $\sim$0.04, far below and thus inconsistent with the obtained data. At $T_{\rm kin}$\,$\sim$\,50\,K,
however, $N$(para-H$_2$CO) $\sim$5$\times$10$^{13}$\,cm$^{-2}$ leads to $T_{\rm mb}$\,(H$_2$CO\,3$_{03}$--2$_{02}$)\,$\sim$\,2.5\,K
and line ratios of order 0.23--0.24. In this case the para-H$_2$CO\,3$_{03}$--2$_{02}$ line is slightly optically thick
($\tau$\,$\sim$\,1.1), while the weaker 3$_{22}$--2$_{21}$ and 3$_{21}$--2$_{20}$ transitions are characterized by $\tau$\,$\sim$\,0.3.
Compared to the optically thin case shown by Fig.\,\ref{fig:Tk-H2CO-ratios}, this leads to a slight overestimate of $T_{\rm kin}$.
Adopting the optically thin scenario, line ratios of 0.25--0.30 lead to kinetic temperatures of
55 to 70\,K instead of the 50\,K adopted in the non-LTE model also accounting for line saturation effects.
Interestingly, even H$_2$CO\,3$_{03}$--2$_{02}$ line opacities of order 5--10 do not lead to larger inaccuracies
(see Fig.\,G.\,2 in \citealt{Immer2016}). While increasing line saturation has a tendency to yield line intensity ratios
closer to unity, this trend is not seen in this case because excitation temperatures of the
H$_2$CO\,3$_{22}$--2$_{21}$ and 3$_{21}$--2$_{20}$ transitions are then significantly
below that of the H$_2$CO\,3$_{03}$--2$_{02}$ transition. To summarize: While our $T_{\rm kin}$ values toward
the cloud cores might be overestimated, they are not grossly overestimated and still clearly favor values well
in excess of the 16\,K derived from ammonia.

In Galactic dense massive star-forming clumps the para-H$_2$CO\,3$_{03}$--2$_{02}$ line tends to be optically
thin \citep{Tang2018b}. Since dust is less prevalent in the LMC than in the Galaxy (e.g. \citealt{Wang2009})
leading to reduced shielding, we may see denser regions with H$_2$CO in the LMC than in the Galaxy, again not supporting
the highly subthermal low kinetic temperature excitation scenario mentioned above. On the other hand, N113 and N159W are the
most intense star-forming regions of the LMC (e.g., \citealt{Wang2009,Lee2016}), which may yield particularly large H$_2$CO column densities.
The peak column densities $N$(para-H$_2$CO) obtained from the para-H$_2$CO\,(3$_{03}$--2$_{02}$) brightness
temperatures of dense H$_2$CO clumps in N113 and N159W (see Table\,\ref{table:Clumps-Parameters}) using the method of \cite{Tang2017a} range
from 0.7 to 9.0$\times$10$^{13}$\,cm$^{-2}$ with an average of $\sim$4.0$\times$10$^{13}$\,cm$^{-2}$ at a density of 10$^5$\,cm$^{-3}$
and from 0.3 to 3.8$\times$10$^{13}$\,cm$^{-2}$ with an average of $\sim$1.4$\times$10$^{13}$\,cm$^{-2}$ at a density of 10$^6$\,cm$^{-3}$,
which are a few times higher than averaged results obtained from entire regions with the APEX telescope \citep{Tang2017b}.

\subsection{Thermal and non-thermal motions}
\label{Sec:Non-thermal motions}
Using the kinetic temperatures derived from the para-H$_2$CO line ratios,
we can derive thermal and non-thermal linewidths
i.e. $\sigma_{\rm T}=\sqrt{\frac{kT_{\rm kin}}{m_{\rm H_2CO}}}$ and
$\sigma_{\rm NT}=\sqrt{\frac{\Delta v^2}{8{\rm ln}2}-\sigma^2_{\rm T}}\thickapprox\Delta v/2.355$,
where $k$ is the Boltzmann constant, $T_{\rm kin}$ is the kinetic
temperature of the gas, $m_{\rm H_2CO}$ is the mass of the formaldehyde
molecule, and $\Delta v$ is the measured FWHM linewidth of para-H$_2$CO\,3$_{03}$--2$_{02}$ \citep{Pan2009}.
Thermal and non-thermal linewidth ranges are listed in Table\,\ref{table:Parameters}.
Thermal and non-thermal linewidths of the individual dense clumps are given in Table\,\ref{table:Clumps-Parameters}.
The thermal linewidth is significantly smaller than the non-thermal linewidth in N113 and N159W,
which indicates that the dense gas traced by para-H$_2$CO is dominated by non-thermal
motions. This agrees with previous measurements derived from para-H$_2$CO\,(3--2 and 4--3) in
the Galactic star-forming region OMC-1 and massive clumps of the Galactic disk \citep{Tang2017a,Tang2018a,Tang2018b}.
Distributions of the non-thermal linewidth in N113 and N159W are shown
in Figs.\,\ref{fig:N113-Tk-H2CO-ratios} and \ref{fig:N159W-Tk-H2CO-ratios}, respectively.
They may suggest that the higher non-thermal linewidths ($\sigma_{\rm NT}$\,>\,1.5\,km\,s$^{-1}$)
associate with the dense clumps and the lower non-thermal linewidths ($\sigma_{\rm NT}$\,$\lesssim$\,1.5)
are located in the outskirts of the dense clumps in our two massive star-forming regions,
indicating that the dense gas traced by para-H$_2$CO is strongly influenced by star-forming activity.

\begin{figure*}[t]
\centering
\includegraphics[width=0.98\textwidth]{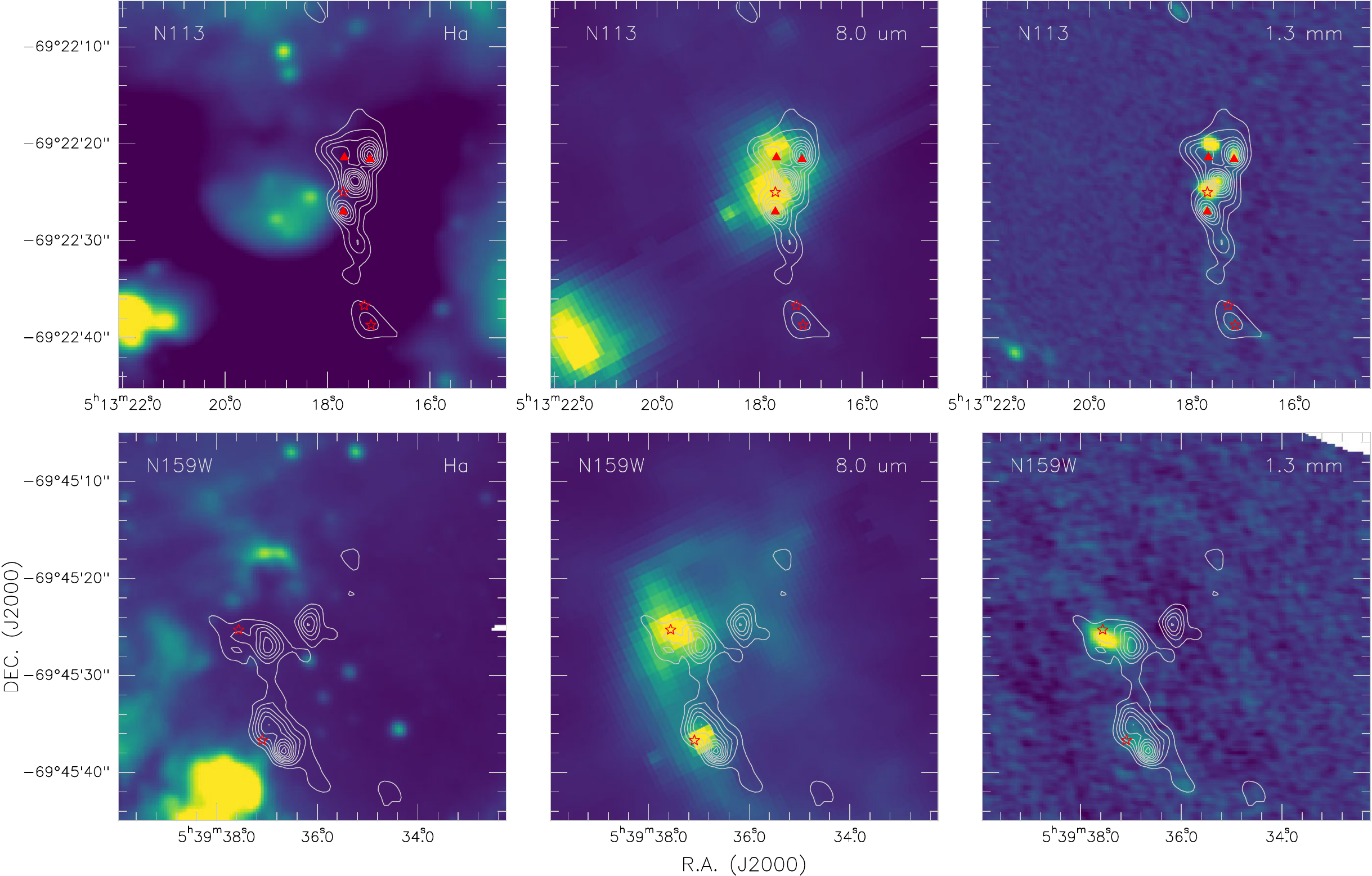}
\caption{Para-H$_2$CO (3$_{03}$--2$_{02}$) with integrated intensity
contours (same as Fig.\,\ref{fig:N113+N159W-maps})
overlaid on H$\alpha$ (\emph{left panels}; \citealt{Smith1998}),
\emph{Spitzer}\,8.0\,$\mu$m (\emph{middle panels}; \citealt{Meixner2006}), and
1.3\,mm continuum (\emph{right panels}; observed with ALMA Band 6;
\citealt{Fukui2015,Sewilo2018}) images of N113 (\emph{top panels})
and N159W (\emph{bottom panels}). Stars and triangles show the locations of
the YSOs (or YSO candidates) and H$_2$O masers \citep{Chen2010,Ellingsen2010,Carlson2012},
respectively.}
\label{fig:N113_N159W_H2CO+Ha+5.8um+1.3mm}
\end{figure*}

Distributions of the Mach number ($\mathcal{M}=\sigma_{\rm NT}/c_{\rm s}$,
where $c_{\rm s}$=$\sqrt{\frac{kT_{\rm kin}}{\mu m_{\rm H}}}$ is
the sound speed, $\mu$\,=\,2.37 is the mean molecular weight for
molecular clouds, and $m_{\rm H}$ is the mass of the hydrogen atom)
in N113 and N159W are shown in Figs.\,\ref{fig:N113-Tk-H2CO-ratios}
and \ref{fig:N159W-Tk-H2CO-ratios}, right panels, respectively.
The Mach numbers of the individual dense clumps in N113 and N159W are listed
in Table\,\ref{table:Clumps-Parameters}. The Mach numbers range from 2.3 to 6.2
with an average of 4.2\,$\pm$\,0.1 in N113 and from 1.6 to 8.3 with an average of
4.5\,$\pm$\,0.1 in N159W (see also Table\,\ref{table:Parameters}). This suggests
that supersonic non-thermal motions (e.g., turbulence, outflows, shocks, and/or magnetic fields)
are dominant in the dense gas traced by H$_2$CO. Locations that are situated
in the outskirts of the clumps of N113 and in the outer regions of the northern clumps
of N159W (see Figs.\,\ref{fig:N113-Tk-H2CO-ratios} and \ref{fig:N159W-Tk-H2CO-ratios}),
have low Mach numbers ($\mathcal{M}$\,$\lesssim$\,3.5). The distribution of Mach numbers
may suggest apparent gradients from southeast to northwest along clumps\,5 and 6 of N159W
(see Fig.\,\ref{fig:N159W-Tk-H2CO-ratios}). Meanwhile, there is a YSO (or YSO candidate) located
in the eastern region of these clumps. This may imply that dense gas probed by para-H$_2$CO may be
influenced by this YSO (or YSO candidate). Higher Mach numbers ($\mathcal{M}$\,>\,3.5)
are associated with the dense clumps in N113 and N159W, again suggesting that dense gas traced
by H$_2$CO is strongly influenced by non-thermal motions (e.g., outflows and shocks)
in these regions. The averaged Mach numbers in N113 and N159W are consistent with previous
observational results in Galactic massive clumps on scales of $\sim$0.1--1.8\,pc
(mean value $\sim$4.2 obtained from para-H$_2$CO\,(3--2 and 4--3); \citealt{Tang2018b}),
but are higher than the values measured with the same transitions of para-H$_2$CO\,(3--2) in OMC-1
on a scale of $\sim$0.06\,pc (mean value $\sim$2.3; \citealt{Tang2018a}).
We have smoothed the OMC-1 para-H$_2$CO\,(3--2) data to a linear resolution of 0.4\,pc.
The derived Mach numbers are 3.0 and 3.5 in the Orion\,KL and Orion South regions, respectively,
which is still slightly lower than the values in N113 and N159W.

One should note that para-H$_2$CO\,(3$_{03}$--2$_{02}$) linewidths in the northern edges of clump 1 and 2 of N113
and in the outskirts of clump 1 and 4 of N159W are narrow <3.0\,km\,s$^{-1}$
(see Figs.\,\ref{fig:N113+N159W-Clump-spectrum}, \ref{fig:N113-Tk-H2CO-ratios}, and \ref{fig:N159W-Tk-H2CO-ratios}),
which is corresponding to only 2--3 channels of our H$_2$CO data. Furthermore, there are two velocity
components around clump 1 of N113 and in the northern region of the clump 4 of N159W
(see Figs.\,\ref{fig:N113-channel-maps} and \ref{fig:N159W-channel-maps}).
For two velocity components the typical linewidths are $\sim$2--3 and $\sim$4--6\,km\,s$^{-1}$
in these regions of N113 and N159W. Our observational set can not
well identify the velocity components with narrow linewidths using Gaussian profiles in these regions.
Therefore, the non-thermal linewidths and the Mach numbers are likely overestimated in these regions.
For the locations with two velocity components in N113 and N159W, the non-thermal linewidths and
the Mach number are weighted with para-H$_2$CO\,(3$_{03}$--2$_{02}$) integrated intensity
in Figs.\,\ref{fig:N113-Tk-H2CO-ratios} and \ref{fig:N159W-Tk-H2CO-ratios}.

\begin{table*}[t]
\tiny
\caption{Gas temperatures at different radius around the young stellar objects or candidates.}
\centering
\begin{tabular}
{cccccccccccccccc}
\hline\hline
&\multicolumn{3}{c}{YSO} &&\multicolumn{2}{c}{$R$\,=\,0.2\,pc}& &\multicolumn{2}{c}{$R$\,=\,0.4\,pc} & &\multicolumn{2}{c}{$R$\,=\,1.0\,pc}\\
\cline{2-4} \cline{6-7} \cline{9-10} \cline{12-13}
Source& R.A.(J2000) & DEC.(J2000) & Luminosity & &$T_{\rm kin}$(Rad) & $T_{\rm kin}$(H$_2$CO) & &$T_{\rm kin}$(Rad) & $T_{\rm kin}$(H$_2$CO) & &$T_{\rm kin}$(Rad) & $T_{\rm kin}$(H$_2$CO)\\
\cline{6-7} \cline{9-10} \cline{12-13}
& $^h$ {} $^m$ {} $^s$ &\degr {} \arcmin {} \arcsec & $\times$10$^5$\,${\rm L}_{\odot}$ &&\multicolumn{2}{c}{K}& &\multicolumn{2}{c}{K} & &\multicolumn{2}{c}{K}\\
\hline
N113    & 05:13:17.28 &--69:22:36.7 &0.16  && 36 & 35 & & 27 & 45 & & 19 & 34 \\
        & 05:13:17.69 &--69:22:25.0 &1.26  && 54 & 48 & & 41 & 57 & & 29 & 53 \\
\hline
N159W   & 05:39:37.09 &--69:45:36.7 &0.82  && 50 & 56 & & 37 & 44 & & 26 & 72 \\
        & 05:39:37.56 &--69:45:25.3 &2.40  && 61 & 69 & & 46 & 54 & & 32 & 41 \\
\hline
\end{tabular}
\label{table:Tkin-YSO}
\tablefoot{The young stellar objects or candidates and their luminosity are taken
from \cite{Chen2010}, \cite{Sewilo2010}, and \cite{Carlson2012}.}
\end{table*}

\section{Discussion}
\label{sect:discussion}
\subsection{Comparison of temperatures obtained from H$_2$CO, CO, and NH$_3$}
\label{sect:Comparison-Tk}
Observations of CO\,$J$\,=\,1--0 to 3--2 indicate that the molecular gas traced by
CO in lower density regions ($n$(H$_2$)\,$<$\,10$^5$\,cm$^{-3}$) may be strongly influenced
by the external FUV emission irradiating giant molecular clouds of the LMC on a scale
of $\sim$10\,pc \citep{Minamidani2008,Minamidani2011}.
The physical and chemical processes in these regions such as formation and destruction
of molecules as well as ionization are dominated by the intense FUV field, which has been
modeled as PDRs (e.g., \citealt{Kaufman1999,Minamidani2011,Lee2016,Lee2019}).
The gas temperature is $T_{\rm kin}$\,>\,150\,K at densities $\sim$10$^3$\,cm$^{-3}$
derived from multitransition data of CO in N159W on a scale of
$\sim$10\,pc \citep{Lee2016}. Our results determined from para-H$_2$CO line ratios
range from 29 to 68\,K with an average of 47.6\,$\pm$\,0.3\,K at density 10$^5$\,cm$^{-3}$
on a scale of 0.4\,pc (see Table\,\ref{table:Parameters}),
which are much lower than values obtained from CO at lower density and even may have to be
slightly modified to the lower side to account for moderate saturation effects
(see Sect.\,\ref{sect:Kinetic-temperature}). This indicates that para-H$_2$CO and CO
trace different kinetic temperature layers in N159W. There is no available
gas temperature derived from multitransition data of CO in N113, so we can not
compare it with our para-H$_2$CO results.

We did not find any evidence for the low $T_{\rm kin}$ value ($\sim$16\,K)
determined from the NH$_3$\,(2,2)/(1,1) line ratio by \cite{Ott2010}.
This is interpreted in terms of a highly embedded extremely obscured dense molecular gas component,
which we do not see in H$_2$CO. A similar situation has also been encountered
in M\,82 \citep{Weiss2001,Muhle2007}, where ammonia suggests $T_{\rm kin}$\,$\sim$\,60\,K,
while H$_2$CO represents gas with temperatures well in excess of 100\,K.
We get moderately high $T_{\rm kin}$ values derived from H$_2$CO in the cores but \cite{Ott2010} obtain a
low $T_{\rm kin}$ value determined from NH$_3$, presumably also from a core.
Ammonia has a particularly low energy threshold for photodissociation
($\sim$4.1\,eV; \citealt{Weiss2001}). Low metallicity environments with low nitrogen abundance
and strong UV radiation (e.g., \citealt{Westerlund1990,Chin1997,Rolleston2002,Wang2009,Ott2010}) in the LMC
may allow ammonia to survive only in the most UV-shielded regions, leading to a low fractional NH$_3$
abundance and a low kinetic temperature of the NH$_3$ emitting gas \citep{Tang2017b}.
It could be explained in the sense, that NH$_3$ is only there, where no protostar has yet heated its environment.
In the outskirts, $T_{\rm kin}$ values from H$_2$CO are low, but CO suggests that there
temperatures are high. This might be explained by the idea that even the H$_2$CO outskirts
are far inside the volume of the CO emitting gas.

\subsection{Radiative heating}
\label{Sec:Radiation-heating}
We compare para-H$_2$CO\,(3$_{03}$--2$_{02}$) integrated intensity
distributions with the H$\alpha$ emission (observed with the UM/CTIO Curtis Schmidt telescope; \citealt{Smith1998}),
\emph{Spitzer}\,8.0\,$\mu$m \citep{Meixner2006}, and 1.3\,mm continuum (observed with ALMA Band 6; \citealt{Fukui2015,Sewilo2018})
emission of N113 and N159W in Fig.\,\ref{fig:N113_N159W_H2CO+Ha+5.8um+1.3mm}.
Obviously, the spatial distributions of para-H$_2$CO\,(3$_{03}$--2$_{02}$)
and H$\alpha$ emission are not well correlated in the two massive star-forming regions.
This suggests that dense gas traced by para-H$_2$CO may only be weakly affected by
the stars exciting the H$\alpha$ emission in N113 and N159W.
Such offsets between molecular and H$\alpha$ emissions are seen in evolved systems (e.g., \citealt{Fukui2010}),
where newly formed massive stars have already dispersed a part of the original molecular gas.
The para-H$_2$CO\,(3$_{03}$--2$_{02}$) integrated intensity distributions agree much better with the \emph{Spitzer}\,8.0\,$\mu$m
emission excited by young massive stars in the N113 and N159W regions (see Fig.\,\ref{fig:N113_N159W_H2CO+Ha+5.8um+1.3mm}).
The para-H$_2$CO\,(3$_{03}$--2$_{02}$) integrated intensity distribution
agrees best with the 1.3\,mm continuum emission in N113 and N159W,
including the dense molecular structure and dust emission peaks
(see Fig.\,\ref{fig:N113_N159W_H2CO+Ha+5.8um+1.3mm}), which implies
that H$_2$CO associates with the dense gas traced by the dust emission.
This is consistent with previous observational results in our Galactic massive
star-forming regions at various evolutionary stages \citep{Immer2014,Tang2017a,Tang2018a,Tang2018b}.

As mentioned, there is no spatial correlation between the dense gas traced by
para-H$_2$CO and the H$\alpha$ distributions in N113 and N159W (see Fig.\,\ref{fig:N113_N159W_H2CO+Ha+5.8um+1.3mm}).
There also appears to be no significant correlation between the gas temperature
structures derived from para-H$_2$CO line ratios and the H$\alpha$ emission distribution
in these two regions (Figs.\,\ref{fig:N113-Tk-H2CO-ratios}, \ref{fig:N159W-Tk-H2CO-ratios},
and \ref{fig:N113_N159W_H2CO+Ha+5.8um+1.3mm}). These findings indicate that dense gas traced
by H$_2$CO is weakly affected by the external FUV radiation in N113 and N159W.
The dense molecular gas ($\gtrsim$\,10$^5$\,cm$^{-3}$) may be effectively shielded
by the lower density molecular gas envelopes in the massive star-forming regions of the LMC.

Previous observations of e.g. H$_2$CO, NH$_3$, CH$_3$CN,
CH$_3$CCH, or CH$_3$OH in Galactic massive star-forming regions
\citep{Lu2014,Giannetti2017,Tang2018a,Tang2018b,Gieser2021} suggest
internal radiative heating of embedded infrared sources of their surrounding dense gas.
Present observations of H$_2$CO in massive star-forming regions of
the LMC imply that the kinetic temperatures traced by para-H$_2$CO\,(3--2) transitions
in dense regions ($\sim$10$^5$\,cm$^{-3}$) are correlated with the ongoing
massive star formation \citep{Tang2017b}. As mentioned in Sect.\,\ref{sect:Kinetic-temperature},
it appears that high gas temperatures obtained from para-H$_2$CO line ratios
associate with dense clumps, YSOs (or YSO candidates), and/or H$_2$O masers in N113 and N159W regions
(see Figs.\,\ref{fig:N113-Tk-H2CO-ratios} and \ref{fig:N159W-Tk-H2CO-ratios}).
This indicates that the warm dense gas traced by para-H$_2$CO is influenced by radiation from
internal embedded infrared sources and/or nearby YSOs in the low-metallicity environment of the LMC.

We determine the gas temperature in regions heated by YSOs following the modified Stefan-Boltzmann
blackbody radiation law, adjusting the emissivity of dust grains smaller than the wavelength at
the characteristic blackbody temperature \citep{Wiseman1998,Tang2018a},
\begin{eqnarray}
\label{equation:radiation}
T_{\rm kin}=2.7\times(\frac{L}{{\rm L}_{\odot}})^{1/5}(\frac{R}{\rm pc})^{-2/5}~{\rm K},
\end{eqnarray}
where the luminosity $L$ is in ${\rm L}_{\odot}$ and the distance $R$ is in parsecs.
Four YSOs (or YSO candidates) with luminosities of 0.16, 0.82, 1.26, and 2.40\,$\times$\,10$^5$\,${\rm L}_{\odot}$
are found in N113 and N159W \citep{Chen2010,Sewilo2010,Carlson2012}.
Assuming these YSOs (or YSO candidates) are the dominant sources at their locations,
the derived gas temperatures at radii of 0.2, 0.4, and 1.0\,pc from Eq.\,(\ref{equation:radiation})
are listed in Table\,\ref{table:Tkin-YSO}. Measured gas temperatures around the four young stellar objects
or candidates derived from para-H$_2$CO line ratios are also listed in Table\,\ref{table:Tkin-YSO},
which agrees well with the results obtained from the radiation law at a radius of 0.2\,pc.
This indicates that dense gas traced by H$_2$CO is significantly influenced by radiation from YSOs (or YSO candidates).
One should note that the distances among YSOs (or YSO candidates), H$_2$O masers, and
1.3\,mm dust emission peaks are approximately $\sim$1\,pc in the N113 and N159W regions
(see Figs.\,\ref{fig:N113-Tk-H2CO-ratios}, \ref{fig:N159W-Tk-H2CO-ratios},
and \ref{fig:N113_N159W_H2CO+Ha+5.8um+1.3mm}). It also shows that the YSO's radiation heating could contribute to
a gas temperature (e.g. $T_{\rm kin}$\,$\sim$\,19\,K for the lowest luminosity YSO (or candidate)
with $L$\,=\,1.6\,$\times$\,10$^4$\,${\rm L}_{\odot}$; see Table\,\ref{table:Tkin-YSO}) even far out,
at a radius of 1\,pc. This implies that the complex temperature structure of the dense gas
in N113 and N159W regions may be also affected by multiple YSOs and/or embedded infrared sources.
For large areas of N113 and N159W the H$\alpha$ radiation regions and the general
stellar radiation field may also contribute to the temperature of the dense gas.

Due to the limited spatial resolution ($\sim$0.4\,pc) of the ALMA data and our limited
2-dimensional perspective, the relation between gas temperature and distance from YSOs
is not perfectly revealed. A follow-up sub-arcsecond angular resolution
(corresponding to $\lesssim$\,0.1\,pc) mapping study of the massive star-forming clump's
temperature structure with H$_2$CO in the LMC will be indispensable and meaningful in the future.

\subsection{Turbulent heating}
\label{Sec:Turbulence}
Turbulent heating seems to be widespread in Galactic massive star-forming
regions on a $\sim$0.1--1.8\,pc scale \citep{Tang2018a,Tang2018b}.
It may play an important role in heating the dense gas in star-forming regions.
Present observations toward Galactic CMZ clouds with para-H$_2$CO\,(3--2 and 4--3)
show that the warm dense gas is heated most likely by turbulence
on a scale of $\sim$1\,pc \citep{Ao2013,Ginsburg2016,Immer2016}.
Observations of, e.g. H$_2$CO, NH$_3$, and CH$_3$CCH in
Galactic star formation regions suggest that the linewidth
is correlated with gas kinetic temperature
\citep{Wouterloot1988,Molinari1996,Jijina1999,Wu2006,Urquhart2011,Urquhart2015,Wienen2012,Lu2014,Tang2017a,Tang2018a,Tang2018b,Giannetti2017}.
Correlations between the kinetic temperature and linewidth are
expected in the case of conversion of turbulent energy into
heat \citep{Gusten1985,Molinari1996,Ginsburg2016}.

\begin{figure}[t]
\vspace*{0.2mm}
\begin{center}
\includegraphics[width=0.48\textwidth]{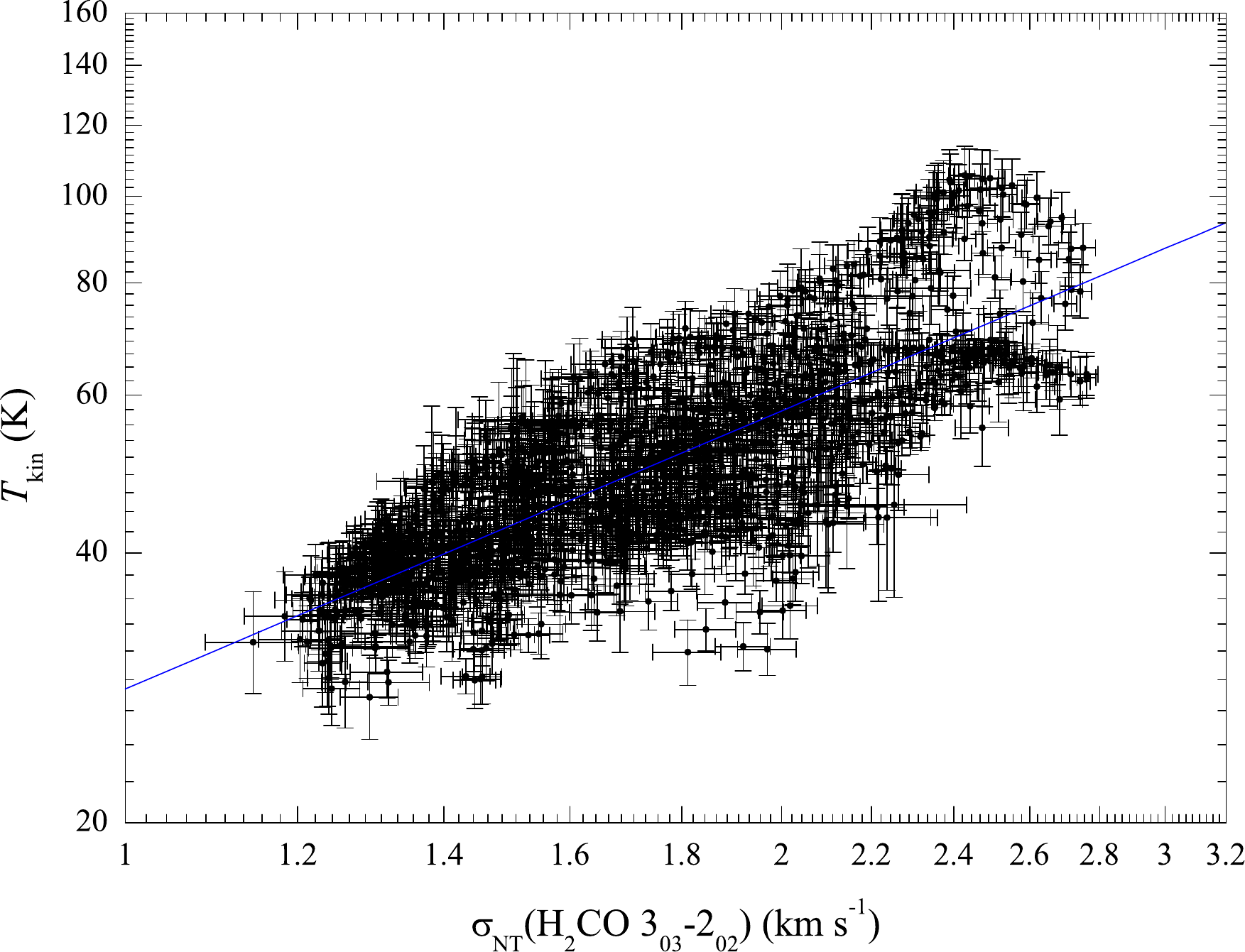}
\end{center}
\caption{Non-thermal velocity dispersion ($\sigma_{\rm NT}$) vs.
gas kinetic temperature derived from the para-H$_2$CO line ratio
for locations (pixels) with the Mach
number $\mathcal{M}$\,$\gtrsim$\,3.5 in N113. The blue line is
the result from an unweighed linear fit, log$T_{\rm kin}=(1.03\pm0.02)\times{\rm log}\sigma_{\rm NT}+(1.45\pm0.01)$,
with a correlation coefficient, $R$, of 0.78.}
\label{fig:Tk-width}
\end{figure}

We examine whether there is a relationship between the turbulence and the
gas temperature derived from the para-H$_2$CO line ratio on a $\sim$0.4\,pc scale in N113.
As discussed in Sect.\,\ref{Sec:Non-thermal motions}, our observational sets
have low channel width ($\sim$1.34\,km\,s$^{-1}$) for para-H$_2$CO data, which strongly affects the estimation
of para-H$_2$CO linewidths in the N159W region, but weakly influences the N113 region which has wider linewidths.
Therefore, we did not examine this relationship in N159W. We adopt the non-thermal velocity dispersion ($\sigma_{\rm NT}$)
of para-H$_2$CO\,(3$_{03}$--2$_{02}$) in good approximation as proxy for the turbulence,
and the kinetic temperature derived from the para-H$_2$CO line ratio as the gas kinetic temperature.
We select positions with strong non-thermal motions ($\mathcal{M}$\,$\gtrsim$\,3.5)
located near the dense clumps (see Figs.\,\ref{fig:N113-Tk-H2CO-ratios} and \ref{fig:N159W-Tk-H2CO-ratios}).
The relation between kinetic temperature and non-thermal velocity dispersion is shown in Fig.\,\ref{fig:Tk-width}.
It indicates that the non-thermal velocity dispersion of para-H$_2$CO is
positively correlated with the gas kinetic temperatures, by a power-law
of the form $T_{\rm kin}\,\propto\,\sigma_{\rm NT}^{1.03\pm0.03}$
with a correlation coefficients $R$ of 0.78, which agrees with results from Galactic
massive star-forming regions ($T_{\rm kin} \propto \sigma_{\rm NT}^{0.66-1.26}$;
gas kinetic temperature measured with para-H$_2$CO\,3--2 and 4--3 line ratios; \citealt{Tang2018a,Tang2018b}).
The intercept and absolute values of the $T_{\rm kin}$--$\sigma_{\rm NT}$(H$_2$CO)
relation in N113 are also consistent with previous results from Galactic massive
star-forming regions on scales of $\sim$0.1--1.8\,pc \citep{Tang2018a,Tang2018b}.
These imply that the higher temperature traced by para-H$_2$CO is related to higher turbulence on a scale of $\sim$0.4\,pc.
To our knowledge, this is the first evidence for dense gas heated in a star-forming region of the LMC by turbulence.
Our $T_{\rm kin}$--$\sigma_{\rm NT}$(H$_2$CO) relation agrees with previous observational results
($T_{\rm kin}\propto\sigma_{\rm NT}^{0.8-1.0}$; gas kinetic temperature measured with NH$_3$ and H$_2$CO)
in molecular clouds of the Galactic center \citep{Gusten1985,Mauersberger1987,Immer2016},
only in terms of slope, not of intercept and absolute value.

We derive the gas kinetic temperature, $T_{\rm turb}$, assuming turbulent heating is the dominant heating process of the
molecular gas in N113 and N159W, following the method applied by \cite{Tang2018a} in their Eq.\,(2),
\begin{eqnarray}
3.3\times10^{-27}~n~\sigma_{\rm NT}^3~L^{-1} = 4\times10^{-33}~n^2~T_{\rm turb}^{1/2}(T_{\rm turb}-T_{\rm dust}) \nonumber \\
+~6\times10^{-29}~n^{1/2}~T_{\rm turb}^3~{\rm d}v/{\rm d}r.
\end{eqnarray}
The gas density $n$ is in units of cm$^{-3}$, the velocity gradient d$v$/d$r$ is in units of km\,s$^{-1}$pc$^{-1}$,
the one-dimensional non-thermal velocity dispersion $\sigma_{\rm NT}$ is in units of km\,s$^{-1}$, and the cloud size $L$ is in units of pc,
and the temperatures $T_{\rm turb}$ and $T_{\rm dust}$ are in K. We assume a cloud size of $\sim$1\,pc (see Fig.\,\ref{fig:N113+N159W-maps}),
a typical velocity gradient of 1\,km\,s$^{-1}$\,pc$^{-1}$, a measured para-H$_2$CO\,(3$_{03}$--2$_{02}$) non-thermal velocity dispersion
(see Table\,\ref{table:Clumps-Parameters}, column 5), and an averaged gas spatial density $n$(H$_2$)\,=\,10$^5$\,cm$^{-3}$ \citep{Tang2017b}.
Previous observations have shown that the gas kinetic temperatures derived from para-H$_2$CO\,(3--2) line ratios
and the dust temperatures obtained from \emph{Herschel} data are in good agreement in the dense star-forming regions of the LMC \citep{Tang2017b},
so here we adopt the gas temperatures obtained from our H$_2$CO line ratios as the dust temperatures.
The derived gas kinetic temperatures $T_{\rm turb}$ are listed in Table\,\ref{table:Clumps-Parameters}.
It seems that the derived $T_{\rm turb}$ values are reasonably well consistent with the gas kinetic temperatures $T_{\rm kin}$(H$_2$CO)
obtained from our H$_2$CO line ratios. This indicates that turbulent heating contributes to the gas temperature on a scale of $\sim$0.4\,pc.
As discussed in Sect.\,\ref{Sec:Radiation-heating}, gas in these dense clumps is also affected by internal star formation activity and/or nearby YSOs.
Therefore, turbulent heating may play an important role in heating the dense gas in the star-forming regions of N113 and N159W.

Observations of the transitions of the para-H$_2$CO\,(3--2) lines with the APEX 12\,m telescope
in the Galactic star-forming region OMC-1\,($\sim$1.1$\times$1.7\,pc$^2$) on a scale
of $\sim$0.06\,pc suggest complex gas heating which is most likely due to a number of
different processes, e.g. star formation activity, radiation, and turbulence \citep{Tang2018a}.
Further observations of para-H$_2$CO\,(3--2 and 4--3) toward dense clumps representing various
evolutionary stages of high-mass star formation in the Galactic plane on scales
of $\sim$0.1--1.8\,pc indicate similar gas heating processes for the dense gas \citep{Tang2018b}.
Our observations reveal also similar gas heating mechanisms in a relatively large area of N113\,($\sim$2.4$\times$6.1\,pc$^2$)
and N159W\,($\sim$3.6$\times$5.1\,pc$^2$) on a scale of $\sim$0.4\,pc in the LMC.
These imply that the gas heating for the dense gas of star-forming regions in low metallicity environments
with strong UV radiation in the LMC may be dominated by internal star formation activity, radiation, and/or turbulence.
The physical and chemical processes of dense gas on a scale of $\sim$0.4\,pc in star-forming regions of
the LMC might be different from those in low density regions ($n$(H$_2$)\,$<$\,10$^5$\,cm$^{-3}$),
so the very dense gas clouds should not be simply modeled as PDRs in low metallicity environments.

\section{Summary}
\label{sect:summary}
We have mapped the kinetic temperature structures of two star formation
regions, N113 and N159W, in the LMC in the
para-H$_2$CO\, $J_{\rm K_aK_c}$\,=\,3$_{03}$--2$_{02}$, 3$_{22}$--2$_{21}$,
and 3$_{21}$--2$_{20}$ lines using ALMA Band 6. The main results are the following:
\begin{enumerate}
\item
There is no significant correlation between the spatial
distributions of the dense gas traced by para-H$_2$CO and H$\alpha$ emission in N113 and N159W.
However, the integrated intensity distributions of para-H$_2$CO are
similar to those of the 1.3\,mm dust emission in N113 and N159W,
indicating that the H$_2$CO emission associates well with dense gas traced
by the emission of the cold dust. There is also some agreement with
the \emph{Spitzer}\,8.0\,$\mu$m emission in N113 and N159W, which
suggests that dense gas traced by para-H$_2$CO associates
with massive star-forming regions.

\item
Using the RADEX non-LTE program, we derive the gas kinetic temperature
by modeling the measured para-H$_2$CO
0.5$\times$[(3$_{22}$--2$_{21}$+3$_{21}$--2$_{20}$)/3$_{03}$--2$_{02}$]
line ratios. The gas kinetic temperatures derived from para-H$_2$CO
line ratios are warm at a spatial density of 10$^5$\,cm$^{-3}$, ranging from
28 to 105\,K in N113 and 29 to 68\,K in N159W. The aforementioned upper
bounds to the kinetic temperature, arising from the cores, may be affected
by moderate line saturation and thus may be slightly overestimated.

\item
The high kinetic temperatures ($T_{\rm kin}$\,$\gtrsim$\,50\,K) of the dense gas traced by
para-H$_2$CO appear to be correlated with the internal embedded
infrared sources and/or YSOs in the N113 and N159W regions.
The lower temperatures ($T_{\rm kin}$\,$<$\,50\,K) are located at the outskirts of the H$_2$CO
distributions of both N113 and N159W. The gas kinetic temperatures show apparent gradients
in some dense H$_2$CO clumps of N113 and N159W.

\item
The kinetic temperatures of the dense gas traced by para-H$_2$CO
appear to be only weakly affected by the external FUV radiation.

\item
The non-thermal velocity dispersions of para-H$_2$CO are correlated
with the gas kinetic temperatures in N113, implying that higher temperatures
traced by para-H$_2$CO are related to turbulence on a $\sim$0.4\,pc scale.

\item
In the LMC, the gas heating for the dense gas may be dominated by internal star formation
activity, radiation, and/or turbulence in star-forming regions with low metallicity environments,
which is consistent with our Galactic massive star-forming regions
located in the Galactic plane.

\end{enumerate}

\begin{acknowledgements}
The authors thank the anonymous referee for helpful comments.
This work acknowledges support by the Heaven Lake Hundred-Talent Program of Xinjiang
Uygur Autonomous Region of China, the National Natural Science Foundation of China
under Grant 11903070, 11433008, and 11973076, the "TianShan Youth Plan" under Grant 2018Q084,
the CAS “Light of West China” Program under Grant Nos.\,2018-XBQNXZ-B-024 and 2020-XBQNXZ-017,
and the Collaborative Research Council 956, subproject A6, funded by the Deutsche
Forschungsgemeinschaft (DFG). C.\,H. acknowledges support by Chinese Academy of
Sciences President's International Fellowship Initiative under Grant No.\,2021VMA0009.
This paper makes use of the following ALMA data:
ADS/JAO.ALMA\#2012.1.00554.S, 2013.1.01136.S, and 2015.1.01388.S.
ALMA is a partnership of ESO (representing its member states),
NSF (USA) and NINS (Japan), together with NRC (Canada), NSC and
ASIAA (Taiwan), and KASI (Republic of Korea), in cooperation
with the Republic of Chile. The Joint ALMA Observatory is
operated by ESO, AUI/NRAO and NAOJ.
This research has used NASA's Astrophysical Data System (ADS).
\end{acknowledgements}

\bibliographystyle{aa}
\bibliography{bibfile}

\begin{appendix}
\onecolumn
\section{H$_2$CO velocity channel maps}

\begin{figure*}[h]
\centering
\includegraphics[width=0.98\textwidth]{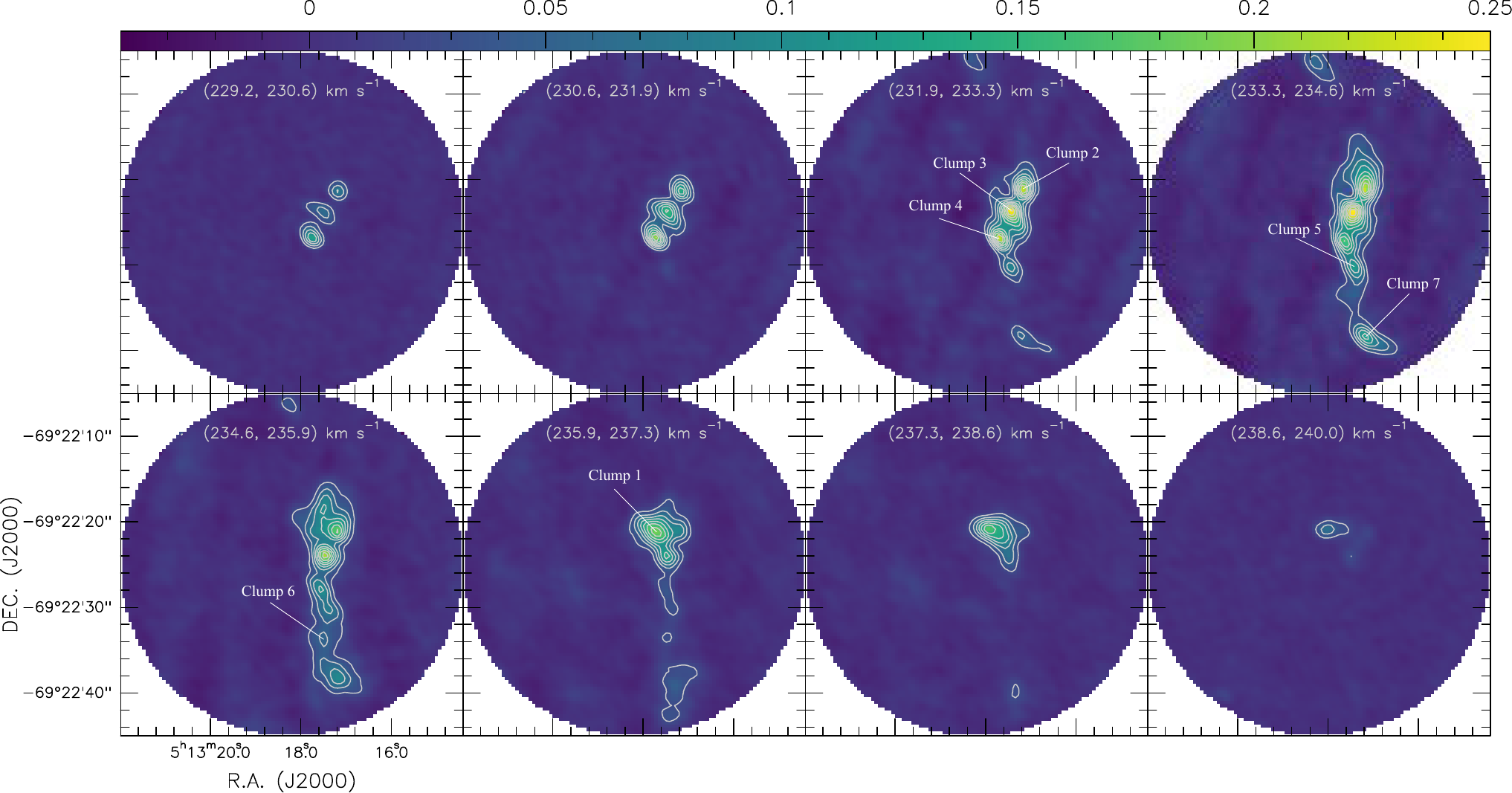}
\caption{Para-H$_2$CO\,3$_{03}$--2$_{02}$ channel maps of N113.
The contour levels are from 10\% to 100\% with steps of 10\%
of the peak intensity (0.25\,Jy\,beam$^{-1}$; color bar in units of Jy\,beam$^{-1}$).
The centers of the fields are given in Fig.\,\ref{fig:N113+N159W-maps}.
Angular scales of 4$''$ correspond to a linear scale of $\sim$1\,pc.
No primary beam correction has been applied.}
\label{fig:N113-channel-maps}
\end{figure*}

\begin{figure*}[h]
\centering
\includegraphics[width=0.98\textwidth]{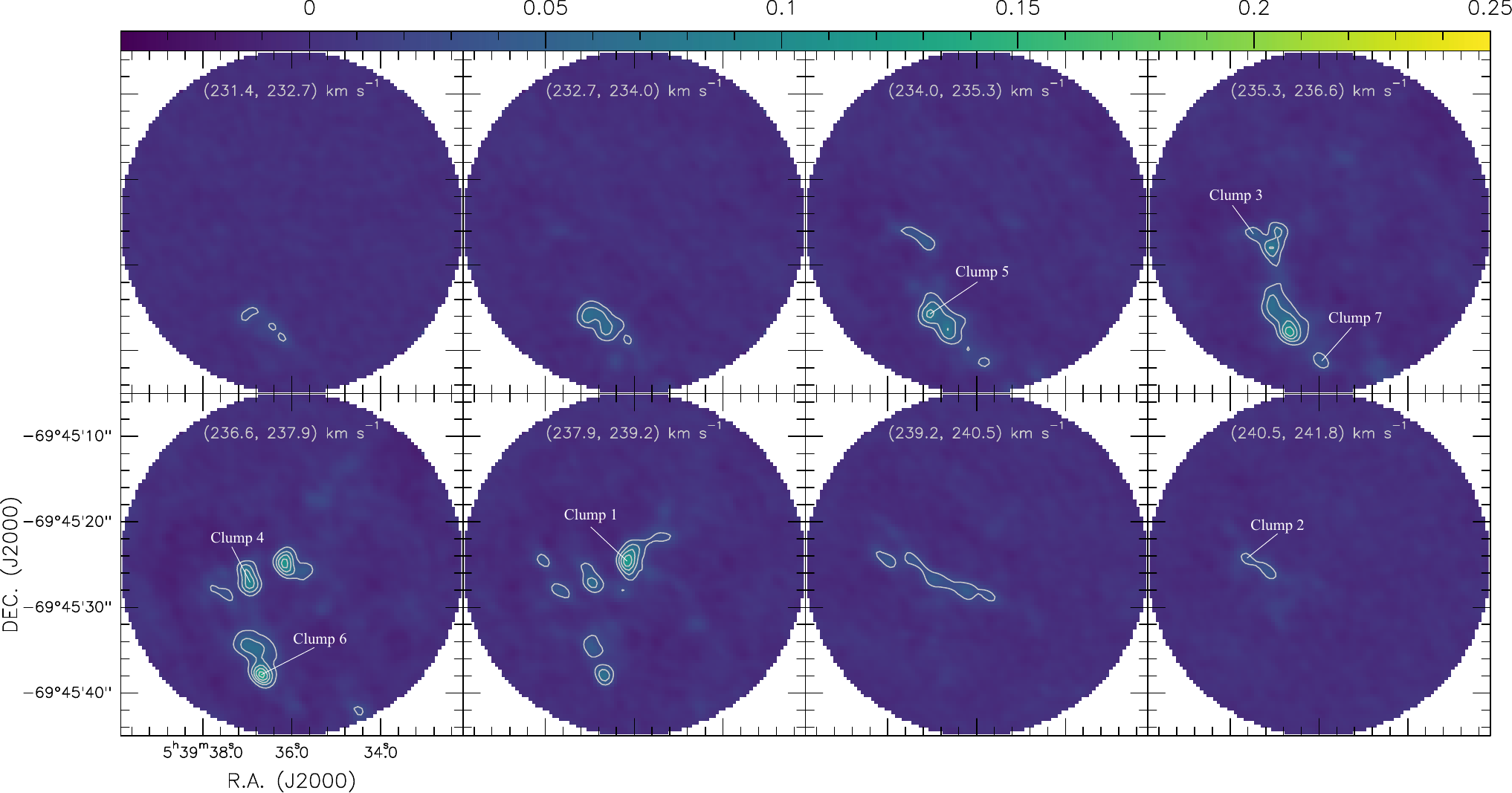}
\caption{Same as Fig.\,\ref{fig:N113-channel-maps}, but for N159W.}
\label{fig:N159W-channel-maps}
\end{figure*}

\end{appendix}

\end{document}